\documentclass[11pt,onecolumn,amssymb,nofootinbib]{revtex4}
\usepackage{amsmath}
\usepackage{amssymb}
\usepackage{graphicx}
\newcommand\beq{\begin{equation}}
\newcommand\eeq{\end{equation}}
\newcommand\beqa{\begin{eqnarray}}
\newcommand\eeqa{\end{eqnarray}}
\newcommand\ro{\hat\rho}
\newcommand\Ho{\hat H}

\newcommand\Qo{\hat Q}

\newcommand\bk{\boldsymbol{k}}
\newcommand\br{\boldsymbol{r}}

\newcommand\bQ{\boldsymbol{Q}}
\newcommand\bu{\boldsymbol{u}}
\newcommand\bv{\boldsymbol{v}}
\newcommand\bqo{\hat{\boldsymbol{q}}}
\newcommand\bro{\hat{\boldsymbol{r}}}
\newcommand\bQo{\hat{\boldsymbol{Q}}}

\begin{document}
\title{CSL reduction rate for rigid bodies}

\author{Luca Ferialdi}
\email{lferialdi@units.it} \affiliation{Department of Physics, University of Trieste, Strada Costiera 11, 34151 Trieste, Italy\\ Istituto Nazionale di Fisica Nucleare, Trieste Section, Via Valerio 2, 34127 Trieste, Italy}

\author{Angelo Bassi}
\email{abassi@units.it} \affiliation{Department of Physics, University of Trieste, Strada Costiera 11, 34151 Trieste, Italy\\ Istituto Nazionale di Fisica Nucleare, Trieste Section, Via Valerio 2, 34127 Trieste, Italy}

\begin{abstract}
In the context of spontaneous wave function collapse models, we investigate the properties of the Continuous Spontaneous Localization (CSL) collapse rate for rigid bodies. By exploiting the Euler-Maclaurin formula, we show that for standard matter the rate for a continuous mass distribution accurately reproduces the exact rate (i.e. the one for a point-like distribution). We compare the exact rate with previous estimates in the literature and we asses their validity. We find that the reduction rate displays a peculiar mass difference effect, which we investigate and describe in detail. We show that the recently proposed layering effect is a consequence of the mass difference effect.
\end{abstract}

\maketitle

\section{Introduction}
Spontaneous collapse models predict a breakdown of the superposition principle in the macroscopic regime, though retaining quantum properties for microscopic systems~\cite{rep1,rep2}. These models are based on a non-linear and stochastic modification of the Schr\"odinger equation which gives very tiny deviations from standard quantum theory for microscopic systems, which become stronger for macroscopic objects, eventually departing from quantum features and recovering classical dynamics.
The most studied collapse model is the mass-proportional CSL model~\cite{CSL}, which is characterized by two parameters: the collapse rate $\lambda$ and the localisation distance $r_C$.
Since the CSL model (like all collapse models) makes different predictions from quantum mechanics, it can be tested against it, allowing to bound its parameters. In recent years, experimental interest increased in this direction and a steady improvement on bounding its parameters has been achieved~\cite{Curetal16,Vinetal16,Biletal16,Pisetal17,Vinetal17,Heletal17,Caretal18,Zheetal20}.

Previous investigations found that the CSL effects on rigid bodies display an important contribution from the geometry of the object~\cite{NimHorHam14,Beletal16,Caretal16}. However, how exactly the CSL collapse rate depends on the geometry of the body and on the superposition distance has never been analyzed in detail. Furthermore, in the literature a continuous mass distribution is often implicitly assumed, but the validity of this assumption has never been investigated. Indeed, since CSL acts on  nucleons, the true mass distribution is point-like.
This paper aims at clarifying these issues, providing a complete analysis of the properties of the CSL collapse rate for rigid bodies. We will mainly work in the position space, denoting space vectors with $\bu$ and $\bv$ in order to avoid confusion with the spatial components of each vector, denoted by $(x,y,z)$. Since calculations in the momentum space are nonetheless instructive, we report them in Appendix A. 
In our estimates we make use of the following values for the CSL parameters: $r_C=10^{-7}$~m and $\lambda=10^{-8}$~s$^{-1}$\cite{Adl07}.

The paper is organized as follows: after reviewing the literature on the CSL collapse rate (Sec. II), we investigate the conditions under which the point-like mass distribution can be replaced by a continuous one (Sec. III). In Section IV we analyse the properties of the collapse rate and we discuss the peculiar \emph{mass difference effect}. In Section V we show that the recently proposed layering effect is a consequence of the mass difference effect, and in Sec. VI we draw our conclusions.

\section{Literature on the CSL collapse rate}

The master equation describing the evolution of the density matrix according to the CSL model in the position space reads~\cite{CSL,rep1,rep2}
\begin{equation}\label{ME}
\frac{d}{dt}\hat{\rho}(t)=-\frac{i}{\hbar}[\Ho,\ro(t)]-\frac{\lambda}{2\pi^{3/2} r_C^3m_N^2}\,\int d^3u\,\int d^3v\,e^{-\frac{(\boldsymbol{u}-\boldsymbol{v})^2}{4r_C^2}} \, [\hat{\mu}(\boldsymbol{u}),[\hat{\mu}(\boldsymbol{v}),\hat{\rho}(t)]]\,,
\end{equation}
where $\Ho$ is the free Hamiltonian, and $r_C$ is a parameter of the model. 
 For a $N$ point-like particles the mass density operator $\hat{\mu}(\boldsymbol{u})$ is
 \beq\label{mu}
\hat{\mu}(\boldsymbol{u})\equiv \sum_i^N m_i\,\delta(\boldsymbol{u}-\hat{\boldsymbol{q}}_i)\,,
\eeq
where $m_i$ is the i-th particle mass and $\hat{\boldsymbol{q}}_i$ its position operator.
The center of mass (c.o.m.) master equation can be obtained by replacing Eq.~\eqref{mu} in Eq.~\eqref{ME}, and rewriting each particle position operator in terms of the c.o.m. ($\bQo$) and relative ($\bro_i$) position operators: $\bqo_i=\bQo+\bro_i$. Under the assumption of rigid body, according to which the relative coordinates are sharply localised (with respect to $r_C$) around the classical positions $\br_i$, i.e. $\langle(\bro_i-\br_i)^2\rangle\ll r_C$, one finds
\begin{equation}\label{MEcm}
\frac{d}{dt}\hat{\rho}_{\text{\tiny{CM}}}(t)=-\frac{i}{\hbar}[\Ho_{\text{\tiny{CM}}},\ro_{\text{\tiny{CM}}}(t)]-\frac{\lambda}{2\pi^{3/2} r_C^3}\,\sum_{i,j=1}^N\,m_im_j\int d^3u\, \left[e^{-\frac{(\bQo+\br_i-\boldsymbol{u})^2}{2r_C^2}},\left[e^{-\frac{(\bQo+\br_j-\boldsymbol{u})^2}{2r_C^2}},\hat{\rho}_{\text{\tiny{CM}}}(t)\right]\right]\,,
\end{equation}
where $\hat{\rho}_{\text{\tiny{CM}}}$ and $\Ho_{\text{\tiny{CM}}}$ denote respectively the density matrix and the Hamiltonian of the center of mass.
While Eq.~\eqref{ME} describes the evolution of the whole body (i.e. of all its particles), Eq.~\eqref{MEcm} describes the evolution of the c.o.m. only.

Most often in experimental situations the displacements $\Delta$ involved are such that $\Delta\ll r_C$. Under this limit, it is possible to expand the master equation~\eqref{MEcm} over $\bQo$ and rewrite it as follows~\cite{AdlBasIpp05}
\begin{equation}\label{MEcmsD}
\frac{d}{dt}\hat{\rho}_{\text{\tiny{CM}}}(t)=-\frac{i}{\hbar}[\Ho_{\text{\tiny{CM}}},\ro_{\text{\tiny{CM}}}(t)]-\sum_{\alpha,\beta}\eta^{\alpha\beta}\left[\Qo_\alpha,\left[\Qo_\beta,\ro_{\text{\tiny{CM}}}(t)\right]\right]\,,
\end{equation}
where $\alpha,\beta=x,y,z$ denote the vector components, and the coefficients $\eta^{\alpha\beta}$ read
\beq\label{etadisc}
\eta^{\alpha\beta}=\frac{\lambda}{8 r_C^4m_N^2}\sum_{i,j=1}^N\,m_im_j\,e^{-\frac{(\br_i-\br_j)^2}{4r_C^2}} \left\{
\begin{array}{rcl}
(r_i^\alpha-r_j^\alpha)\, (r_i^\beta-r_j^\beta)&\,\,&\alpha\neq\beta\\
2r_C^2-(r_i^\alpha-r_j^\alpha)^2&\,\,&\alpha=\beta
\end{array}\right.\,.
\eeq
These diffusion coefficients are determined by the geometry of the body, and for simple geometries can be computed exactly~\cite{NimHorHam14,Beletal16,Caretal16}. In the literature, the diffusion coefficients are most often estimated in the momentum space~\cite{NimHorHam14,Caretal16,Vinetal17,CarFerBas18} (see Appendix A).

Coming back to Eq.~\eqref{ME}, we introduce the vector $|q^{\text{\tiny{L}}}\rangle$, where $q^{\text{\tiny{L}}}=\{\boldsymbol{q}^{\text{\tiny{L}}}_i\}_{i=1}^N$ is the set of particles positions $\boldsymbol{q}^{\text{\tiny{L}}}_i=(x^{\text{\tiny{L}}}_i,y^{\text{\tiny{L}}}_i,z^{\text{\tiny{L}}}_i)$. Since we are interested to the collapse properties of the body, we neglect the free evolution, and we take the matrix element $\langle q^{\text{\tiny{L}}}|\cdot|q^{\text{\tiny{R}}}\rangle$ obtaining
\begin{equation}
\langle q^{\text{\tiny{L}}}|\hat{\rho}(t)|q^{\text{\tiny{R}}}\rangle=e^{-\Gamma(q^{\text{\tiny{L}}},q^{\text{\tiny{R}}})\,t}\,\langle q^{\text{\tiny{L}}}|\hat{\rho}(0)|q^{\text{\tiny{R}}}\rangle\,,
\end{equation}
with
\beqa\label{rate}
\Gamma(q^{\text{\tiny{L}}},q^{\text{\tiny{R}}})&=&\frac{\lambda}{2m_N^2}\,\int d^3u\,\int d^3v\,e^{-\frac{(\boldsymbol{u}-\boldsymbol{v})^2}{4r_C^2}} \Big(\mu^{\text{\tiny{L}}}(\bu)-\mu^{\text{\tiny{R}}}(\bu)\Big)\Big(\mu^{\text{\tiny{L}}}(\bv)-\mu^{\text{\tiny{R}}}(\bv)\Big)\,,
\eeqa
where $\mu^{\text{\tiny{L}}}(\bu)\equiv\sum_i^N m_i\,\delta(\boldsymbol{u}-\boldsymbol{q}_i^{\text{\tiny{L}}})$ (similar definition holds for $\mu^{\text{\tiny{R}}}(\bu)$).
We thus see that the collapse rate ~$\Gamma(q^{\text{\tiny{L}}},q^{\text{\tiny{R}}})$ depends on the specific mass distributions  $\mu^{\text{\tiny{L}}}(\bu)$, $\mu^{\text{\tiny{R}}}(\bu)$, and  in general needs to be computed case by case. In the following we will drop the explicit dependence on the sets of positions $q^{\text{\tiny{L}}},q^{\text{\tiny{R}}}$ and denote the reduction rate simply by $\Gamma$. We will further assume the vector $|q^{\text{\tiny{R}}}\rangle$ to be a rigid displacement of $|q^{\text{\tiny{L}}}\rangle$ by a vector $\boldsymbol{\Delta}$, i.e. that the i-th position of the state $|q^{\text{\tiny{R}}}\rangle$ is $\boldsymbol{q}^{\text{\tiny{R}}}_i=\boldsymbol{q}^{\text{\tiny{L}}}_i+\boldsymbol{\Delta}$. In order to simplify the treatment we will consider a displacement oriented in the z direction: $\boldsymbol{\Delta}=(0,0,\Delta)$. The extension of our results to a general $\boldsymbol{\Delta}$ is straightforward. Furthermore, one can show that the total reduction rate defined in Eq.~\eqref{rate} coincides with the c.o.m. reduction rate: taking the matrix element $\langle\bQ^{\text{\tiny{L}}}|\cdot|\bQ^{\text{\tiny{R}}}\rangle$ of the c.o.m. master equation~\eqref{MEcm} (where $|\bQ^{\text{\tiny{R}}}\rangle$ is a rigid displacement by an amount $\boldsymbol{\Delta}$ of $|\bQ^{\text{\tiny{L}}}\rangle$) leads to the desired result. 
Equation~\eqref{rate} is the main formula of the paper, and it will be used to calculate the reduction rate of rigid bodies for different mass distributions. For the following discussion, it is important to stress that the double integral in Eq.~\eqref{rate} \emph{measures the correlation of the difference of the mass distributions} over a Gaussian distribution with spread $\sqrt{2}r_C$. This feature is a direct consequence of the double commutator displayed by the master equation~\eqref{ME} and, as we will see, plays a crucial role in defining the properties of the collapse rate.

A first estimate of the c.o.m. collapse rate was provided by Ghirardi, Pearle and Rimini~\cite{CSL}, who considered a homogeneous mass density distribution in the limit $r_C\rightarrow0$, that corresponds to an extremely sharped localisation Gaussian (essentially a Dirac delta). They found
\beq\label{gpr}
\Gamma_{\text{\tiny{GPR}}}=6\,\sqrt{\pi}\,\lambda \,n\, N_{\text{\tiny{OUT}}}\,
\eeq
where $n$ is the number of nucleons contained by a sphere of radius $r_C$. $N_{\text{\tiny{OUT}}}$ is the number of nucleons in the volume of the body in a state $|\bQ^{\text{\tiny{L}}}\rangle$ that do not lie in the volume when the state is $|\bQ^{\text{\tiny{R}}}\rangle$, thus implying that the rate depends linearly on the displacement $\Delta$. We remark that the limit $r_C\rightarrow0$ essentially coincides with the requirement that both $\Delta$ and the body dimension must be much larger than $r_C$.

Later, Adler considered a body with discrete mass distribution and a displacement $\Delta$ such that the states do not overlap (``large superposition'').
He showed that the nucleons in a volume of size $\ll r_C$ contribute quadratically to the rate, while volumes distant $\gg r_C$ from each other contribute linearly~\cite{Adl07}. He then ideally divides the considered body in $N$ spheres of radius $r_C$, each containing $n$ nucleons ($N_{\text{\tiny{TOT}}}=n\,N$ being the total number of nucleons in the body), and evaluates the collapse rate as
\beq\label{adler}
\Gamma_{\text{\tiny{A}}}= \lambda\, n^2N\left\{
\begin{array}{rcl}
\frac{\Delta^2}{2r_C^2}&\,\mathrm{if}\,&\Delta\ll r_C \\
1&\,\mathrm{if}\,&\Delta\gg r_C
\end{array}\right.\,.
\eeq
We remark that this idealized division of the body ($N$ spheres of radius $r_C$ close to each other) does not meet Adler's assumptions (volumes of size $\ll r_C$, distant from each other $\gg r_C$), thus Eq.~\eqref{adler} should be considered as an estimate of the rate. 
We thus see that there are some differences between $\Gamma_{\text{\tiny{GPR}}}$ and $\Gamma_{\text{\tiny{A}}}$: the first depends linearly on $\Delta$, while the latter does not (in the limit of large displacement where $\Gamma_{\text{\tiny{GPR}}}$ is defined); also $\Gamma_{\text{\tiny{GPR}}}$ takes into account the mass difference of the two states onto which the rate is evaluated, while $\Gamma_{\text{\tiny{A}}}$ is valid only for large superpositions (for which the mass difference coincides with the total mass of the body). As we will see both these features play an important role. The regimes of validity of $\Gamma_{\text{\tiny{GPR}}}$ and $\Gamma_{\text{\tiny{A}}}$ are summarized in Table~I.
\begin{table}\label{table}
\begin{center}
  \begin{tabular}{c|c|c|c|}
  &$R$ vs $r_C$&$\Delta$ vs $r_C$& $R$ vs $\Delta$\\
  \hline
$\Gamma_{\text{\tiny{GPR}}}$&$R\gg r_C$& $ \Delta\gg r_C$& any $R/\Delta$\\
\hline
$\Gamma_{\text{\tiny{A}}}$&any $R/r_C$ &  $\Delta\gg r_C$ or $\Delta\ll r_C$&$\Delta>2R$\\
\hline
\end{tabular}
\end{center}
 \caption{Regimes of validity of $\Gamma_{\text{\tiny{GPR}}}$ and $\Gamma_{\text{\tiny{A}}}$ in terms of body size ($R$), superposition distance ($\Delta$) and localisation distance ($r_C$). $\Gamma_{\text{\tiny{GPR}}}$ is defined for $R$ and $\Delta$ both much larger than $r_C$, and for any ratio between $R$ and $\Delta$. $\Gamma_{\text{\tiny{A}}}$ is defined for $\Delta$ much larger or much smaller than $r_C$, and for any $R$, provided that $\Delta>2R$. 
 }
\end{table}

\section{Discrete vs Continuous mass distribution}
The mass of a body is mainly concentrated in the nuclei, so its mass density is a discrete distribution of spheres of nuclear size. Since nuclei are extremely small with respect to other distances involved in the collapse process, the mass distribution can be considered as point-like, thus explaining the definition in Eq.~\eqref{mu}. Reminding that $\boldsymbol{q}^{\text{\tiny{R}}}_i=\boldsymbol{q}^{\text{\tiny{L}}}_i+\boldsymbol{\Delta}$, one finds that the collapse rate of Eq.~\eqref{rate} for such a point-like mass distribution becomes
\beq\label{gammaD}
\Gamma_\text{\tiny{D}}=\frac{\lambda}{m_N^2}\,\sum_{i,j=1}^N\, m_im_j\,\left(e^{-\frac{(\boldsymbol{q}_i-\boldsymbol{q}_j)^2}{4r_C^2}}-e^{-\frac{(\boldsymbol{q}_i-\boldsymbol{q}_j-\boldsymbol{\Delta})^2}{4r_C^2}}\right)\,,
\eeq
where we have dropped the superscript $^\text{\tiny{L}}$ for notational convenience.
In the literature the collapse rate is often calculated by relaxing the definition~\eqref{mu} and by implicitly assuming a continuous mass distribution~\cite{NimHorHam14,Beletal16,Caretal16}, although the range of validity of this assumption has never been investigated. For a continuous mass distribution Eq.~\eqref{rate} becomes
\beq\label{gammaC}
\Gamma_\text{\tiny{C}}=\frac{\lambda}{m_N^2}\int d^3u\int d^3v\, \mu(\boldsymbol{u})\mu(\boldsymbol{v})\left(e^{-\frac{(\boldsymbol{u}-\boldsymbol{v})^2}{4r_C^2}}-e^{-\frac{(\boldsymbol{u}-\boldsymbol{v}-\boldsymbol{\Delta})^2}{4r_C^2}}\right)\,,
\eeq
which in the limit of small displacement $\Delta\ll r_C$ reduces to
\beq\label{gammasmall}
\Gamma_\text{\tiny{C}}=\frac{\lambda}{m_N^2}\frac{\Delta^2}{4r_C^2}\int d^3u\int d^3v\,\mu(\boldsymbol{u}) \mu(\boldsymbol{v}) \left[1-\frac{(u_z-v_z)^2}{2r_C^2}\right] e^{-\frac{(\boldsymbol{u}-\boldsymbol{v})^2}{4r_C^2}}\equiv\Delta^2\eta^{zz}\,,
\eeq
where $\eta^{zz}$ is simply the continuous version of Eq.~\eqref{etadisc} for $\alpha=\beta=z$. Obviously, if one replaces the discrete mass distribution $\mu(\boldsymbol{u})= \sum_im_i\,\delta(\boldsymbol{u}-\boldsymbol{q}_i)$ in Eq.~\eqref{gammaC} one recovers Eq.~\eqref{gammaD}.

In order to investigate the validity of the ``continuous mass density'' assumption we start from $\Gamma_\text{\tiny{D}}$ and we consider a cuboidal body of sides $L_x$, $L_y$, $L_z$, which we model as a cubic crystal of lattice constant $l$, with $N_{\text{\tiny{S}}}$ sites each having an atom with $n_A$ nucleons. This geometry is particularly convenient for two reasons: it allows to simplify significantly the sums in Eq.~\eqref{gammaD} (the square distance between two sites is always a multiple of $l^2$); and, more importantly, it allows to exploit the Euler-Maclaurin (EM) formula~\cite{euler,maclaurin,AbrSte72,apostol} to estimate the error that is made when approximating a discrete sum by an integral. For a generic continuous function $f(x)$ with $(2p+1)$-th continuous derivative, the EM formula reads
\beq\label{EM}
\sum_{i=1}^Nf(i)=\int_0^Ndx f(x)+\frac{1}{2}\left[f(N)-f(0)\right]+\sum_{k=1}^{p}\frac{B_{2k}}{2k!}\left[f^{(2k-1)}(N)-f^{(2k-1)}(0)\right]+R_p\,,
\eeq
where 
$B_k$ is the $k$-th Bernoulli number, and $f^{(k)}$ is the $k$-th derivative of $f(x)$. The value of $p$ sets the order of approximation of the error estimate, and can be chosen in such a way to minimize the remainder $R_p$. 
We refer the reader to Appendix B for a more mathematical statement of the formula. In order to apply this formula to $\Gamma_\text{\tiny{D}}$, we decompose the double sum of Eq.~\eqref{gammaD} in a product of three double sums, two for each spatial direction. Equation~\eqref{gammaD} can thus be rewritten as follows
\beq\label{gammaDser}
\Gamma_\text{\tiny{D}}=\lambda n_A^2\, \left(\sum_{i_x,j_x=1}^{N_x}\,e^{-\frac{l^2(i_x-j_x)^2}{4r_C^2}}\right)\left(\sum_{i_y,j_y=1}^{N_y}\,e^{-\frac{l^2(i_y-j_y)^2}{4r_C^2}}\right)\left(\sum_{i_z,j_z=1}^{N_z}\,e^{-\frac{l^2(i_z-j_z)^2}{4r_C^2}}-e^{-\frac{l^2(i_z-j_z-\delta)^2}{4r_C^2}}\right)\,,
\eeq
where $\delta=\Delta/l$, $N_\alpha=L_\alpha/l$ ($\alpha=x,y,z$) is the number of sites in each direction, and $\prod_\alpha N_\alpha=N_{\text{\tiny{S}}}$. Although the original EM formula involves single sums, it is possible to extend it to double sums, as shown in Appendix B. At lowest order of the EM formula, one can show that the second term of the double sum in the z direction can be approximated by a double integral as follows
\beq\label{sumZ}
\sum_{i_z,j_z=1}^{N_z}\,e^{-\frac{l^2(i_z-j_z-\delta)^2}{4r_C^2}}= N_z^2\,g_{\text{\tiny{$\Delta$}}}(L_z)+\frac{1}{3}\left(e^{-\frac{\Delta^2}{4r_C^2}}-\frac{1}{2}e^{-\frac{(L_z-\Delta)^2}{4r_C^2}}-\frac{1}{2}e^{-\frac{(L_z+\Delta)^2}{4r_C^2}}\right)+O\left(\frac{l^2}{2r_C^2}\right)\,,
\eeq
where we have introduced the function
\beq\label{gD}
g_{\text{\tiny{$\Delta$}}}(L_z)\equiv \frac{1}{L_z^2}\int_0^{L_z}du_z\int_0^{L_z}dv_z\,e^{-\frac{(u_z-v_z-\Delta)^2}{4r_C^2}}\,.
\eeq 
The term $O(l^2/2r_C^2)$ denotes the fact that the contributions coming from Eq.~\eqref{EM} not displayed in Eq.~\eqref{sumZ} are at least of the order $l^2/2r_C^2$ (see Appendix B for a more detailed discussion).
Working with a cuboid has also the advantage that Eq.~\eqref{gD} can be integrated exactly, leading to
 \beq\label{gDelta}
g_{\text{\tiny{$\Delta$}}}(L_z)=\frac{1}{2}g(L_z-\Delta)+\frac{1}{2}g(L_z+\Delta)-g(\Delta)\,,
\eeq
with
 \beq\label{gfac}
g(x)\equiv g_{\text{\tiny{$\Delta=0$}}}(x)=\frac{4r_C^2}{x^2}\left(e^{-\frac{x^2}{4r_C^2}}-1+\sqrt{\pi}\frac{x}{2r_C}\mathrm{erf}\left[\frac{x}{2r_C}\right]\right)\,.
\eeq
All other double sums contributing to $\Gamma_D$ can be rewritten as double integrals simply by setting $\Delta=0$ in Eq.~\eqref{sumZ}. We replace Eq.~\eqref{sumZ}, together with the similar expressions obtained for the series in the $x$ and $y$ directions, in Eq.~\eqref{gammaDser} obtaining
\beq\label{DC}
\Gamma_\text{\tiny{D}}=\lambda \,N_{\text{\tiny{TOT}}}^2\, g(L_x)g(L_y)\left[g(L_z)-g_{\text{\tiny{$\Delta$}}}(L_z)\right]+\mathcal{E}\,,
\eeq
where $N_{\text{\tiny{TOT}}}=n_A\,N_{\text{\tiny{S}}}$ is the total number of nucleons in the body, and $\mathcal{E}$ collects all the remaining terms of the product of the three double series. 
One can easily recognize that the first term of the right hand side of Eq.~\eqref{DC} is nothing but Eq.~\eqref{gammaC} for a cuboidal homogeneous mass distribution with density $\varrho=N_{\text{\tiny{TOT}}}\,m_N/(L_xL_yL_z)$, i.e.
\beq\label{gammacuboid}
\Gamma_\text{\tiny{C}}=\lambda \,N_{\text{\tiny{TOT}}}^2\, g(L_x)g(L_y)\left[g(L_z)-g_{\text{\tiny{$\Delta$}}}(L_z)\right]\,.
\eeq
We can thus rewrite Eq.~\eqref{DC} as follows
\beq\label{gammaDvsC}
\Gamma_\text{\tiny{D}}=\Gamma_\text{\tiny{C}}+\mathcal{E}\,,
\eeq
where $\mathcal{E}$ can be understood as the error made when approximating $\Gamma_\text{\tiny{D}}$ with $\Gamma_\text{\tiny{C}}$. In order to measure how good such approximation is, 
we introduce the relative error $\mathcal{E}^R=\mathcal{E}/\Gamma_\text{\tiny{C}}$ and we estimate it for two different experimental scenarios: $\Delta\ll r_C$ and $L_\alpha\gg r_C$, relevant for non-interferometric experiments (typical resonator size is of the order of $10^{-5}$ m~\cite{Vinetal17}); $\Delta\ll r_C$ and $L_\alpha\ll r_C$, relevant for molecular interferometry (typical macromolecule size is of the order of $10^{-9}$~m~\cite{molecules}).
In the first case, one can show that the leading contribution to the error is
\beq
\mathcal{E}= N_x^2\,g(L_x)\,N_y^2\,g(L_y)\frac{\Delta^2}{12\,r_C^2}+O\left(\frac{L_x\,r_C\,\Delta^2}{l^4}\right)\,,
\eeq
where for simplicity we have set $L_y=L_x$, and the relative error is
\beq
\mathcal{E}^R=\frac{l^2}{6\,r_C^2}+O\left(\frac{l^2}{r_C\,L_x}\right)\,.
\eeq
We thus see that the accuracy of the approximation~\eqref{gammaDvsC}  depends only on the ratio $l/r_C$. 
A similar result is obtained in the case $L_\alpha\ll r_C$: the leading contribution to the error is
\beq
\mathcal{E}=\frac{L_x^2}{6\,r_C^2}N_x^2\,g(L_x)\,N_z^2\,\left[g(L_z)-g_{\text{\tiny{$\Delta$}}}(L_z)\right]+O\left(\frac{L_x^2\,L_y^2\,L_z^2\,\Delta^2}{r_C^6\,l^2}\right)\,,
\eeq
and the relative error becomes
\beq
\mathcal{E}^R=\frac{l^2}{3\, r_C^2}+O\left(\frac{l^4}{r_C^4}\right)\,.
\eeq
We thus see that also when the number of nucleons involved is very small, the accuracy of the description given by $\Gamma_\text{\tiny{C}}$ depends only on the ratio $l/r_C$. Accordingly, $\Gamma_\text{\tiny{C}}$ accurately approximates $\Gamma_\text{\tiny{D}}$ for any $l\lesssim\sqrt{2} r_C$, i.e. whenever the last term of Eq.~\eqref{sumZ} is negligible. For a standard piece of matter with $l=10^{-10}$ m,  $\Gamma_\text{\tiny{C}}$ gives an extremely precise description of the exact reduction rate.
However, there are experimental situations, e.g. with cold atoms~\cite{coldat}, where the average distance among particles can be larger than $\sqrt{2} r_C$.
As previously mentioned, Eq.~\eqref{sumZ} clearly shows that in this case, the description provided by $\Gamma_\text{\tiny{C}}$ is not accurate, and one needs to compute $\Gamma_\text{\tiny{D}}$.

Physically, we can understand the fact that $\Gamma_\text{\tiny{C}}$ accurately approximates $\Gamma_\text{\tiny{D}}$ only for $l\lesssim\sqrt{2} r_C$ as follows: The reduction rate $\Gamma_\text{\tiny{D}}$ in its form~\eqref{gammaDser} can be understood as if there is a Gaussian function of width $\sqrt{2}r_C$ sitting at each particle's position. The discrete mass distribution is thus ``spreaded'' by the Gaussians over a distance $\sqrt{2}r_C$. Whenever the distance among the particles is smaller than $\sqrt{2}r_C$, the Gaussian significantly overlap and the mass distribution effectively results continuous. When $l>\sqrt{2}r_C$ the Gaussians essentially do not overlap, and the continuous picture fails. In the light of the above results, in what follows we will consider $\Gamma_\text{\tiny{C}}$ for evaluating the collapse rate.

\section{Properties of the CSL collapse rate}

In this section we investigate in detail the features of the reduction rate for bodies with cuboidal geometry.
It is clear from the definiton~\eqref{rate} that the collapse rate depends on the difference of the mass densities of the two states onto which the rate is evaluated. However, how this practically influences the behaviour of the reduction rate has not been investigated.
We start our analysis by considering a homogeneous cubic body of side $L$, and we compare the exact reduction rate $\Gamma_\text{\tiny{C}}$ of Eq.~\eqref{gammacuboid} with the estimates $\Gamma_\text{\tiny{GPR}}$ and $\Gamma_\text{\tiny{A}}$ of Eqs.~\eqref{gpr},~\eqref{adler} respectively. 
\begin{figure}
 \begin{minipage}[h]{0.4\textwidth}
     \includegraphics[width=\textwidth]{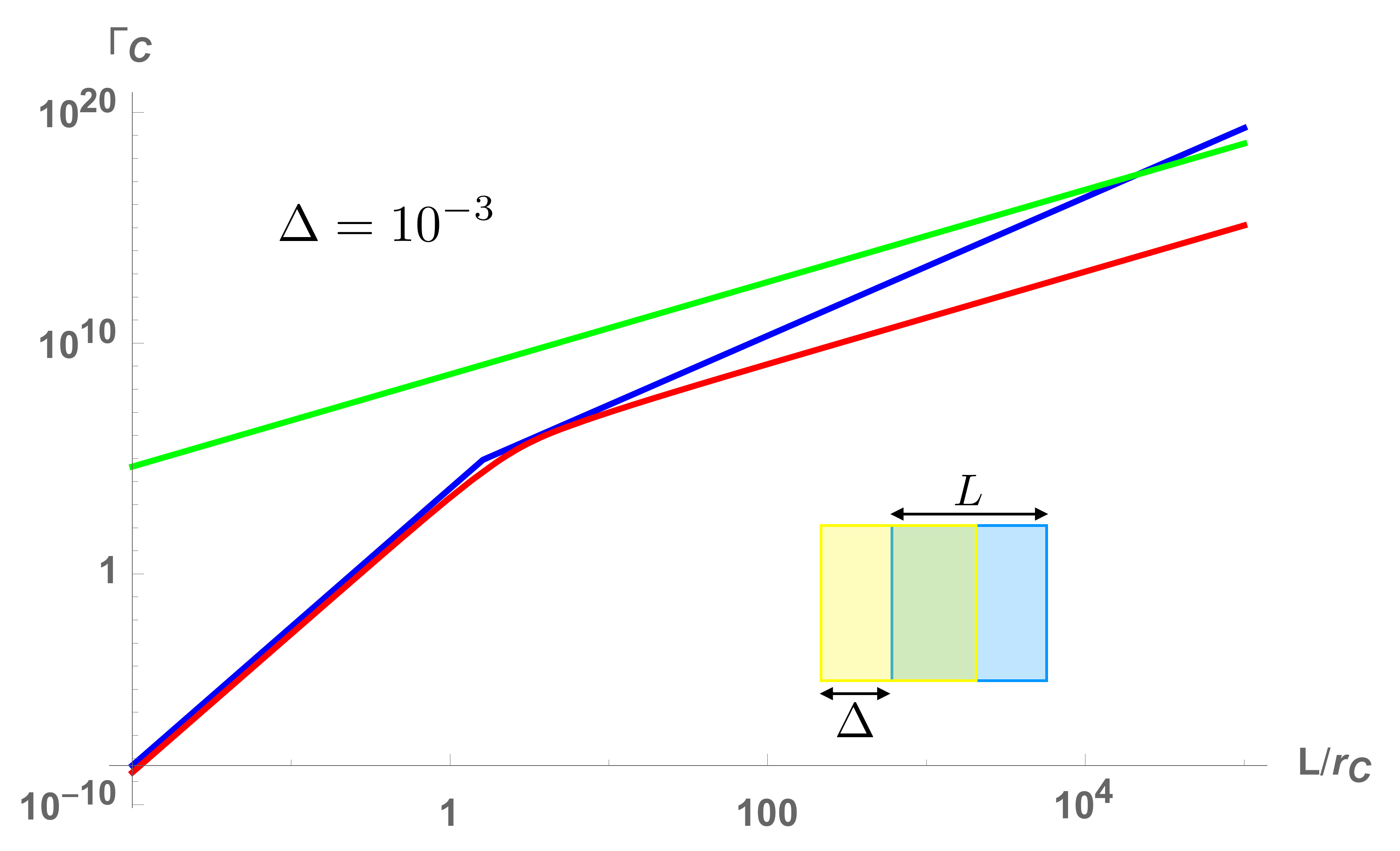} 
  \end{minipage}
  \hspace{1cm}
  \begin{minipage}[h]{0.4\textwidth}
     \includegraphics[width=\textwidth]{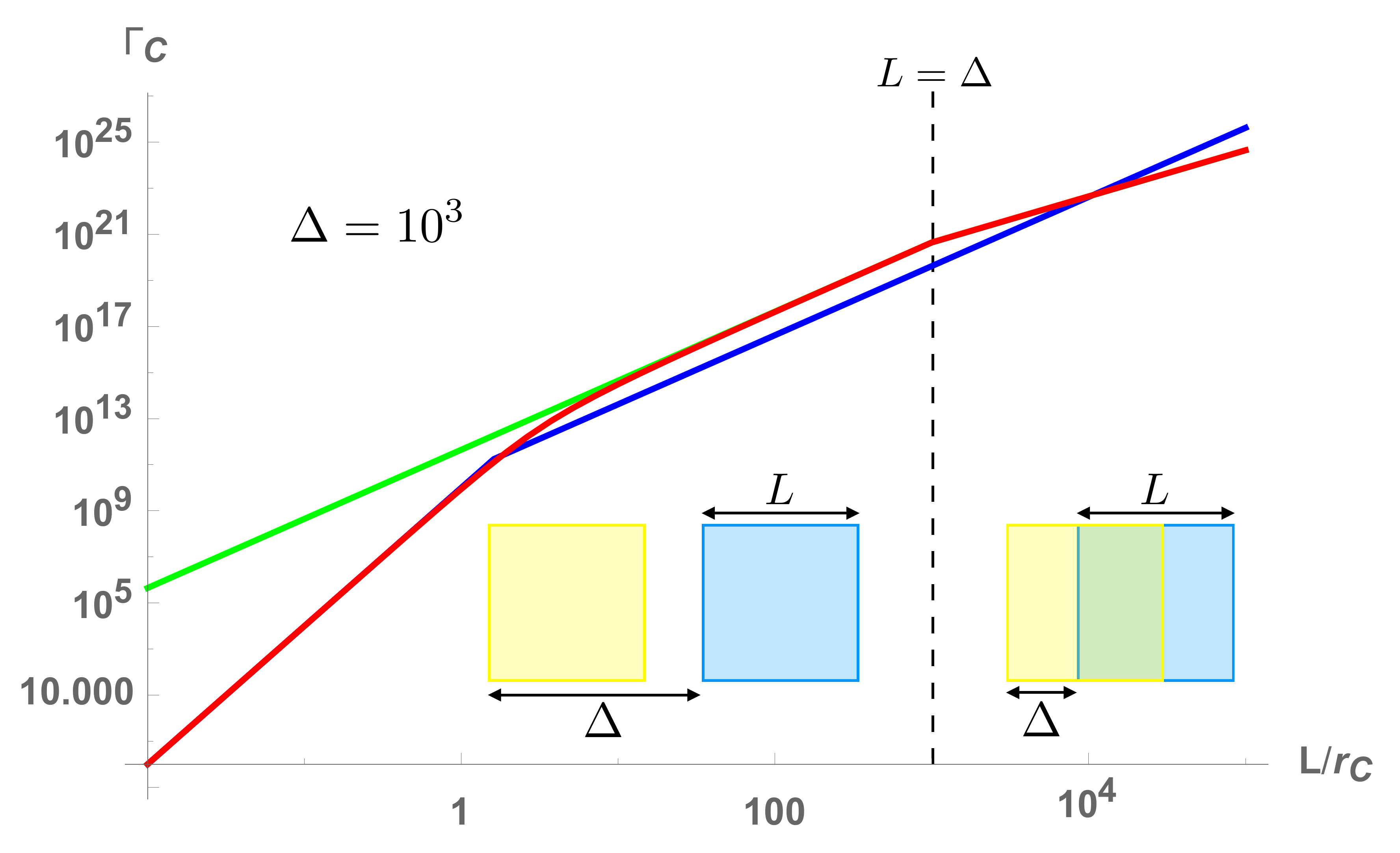}
  \end{minipage}
   \caption{Log-Log plot of the reduction rate of a cube as a function of the side $L$ in units of $r_C$, for $\Delta=10^{-3}\, r_C$ (left panel) and $\Delta=10^{3}\,r_C$ (right panel). The lines correspond to: exact reduction rate $\Gamma_\text{\tiny{C}}$ of Eq.~\eqref{gammacuboid} (red line), $\Gamma_\text{\tiny{GPR}}$ of Eq.~\eqref{gpr} (green line) and $\Gamma_\text{\tiny{A}}$ of Eq.~\eqref{adler} (blue line). Body density is $10^{30}$ nucleons/m$^3$, $r_C=10^{-7}$ m, and $\lambda=10^{-8}$ s$^{-1}$. Yellow square represents the state $|q^{\text{\tiny{L}}}\rangle$, blue square represents  $|q^{\text{\tiny{R}}}\rangle$. In the left panel, the range of $L$ is such that the two states always overlap. In the right panel, the dashed line separates the region where the states do not overlap (left), from the region where they do so (right). In both panels, the change of slope of $\Gamma_\text{\tiny{C}}$ (red line) around $L\simeq\sqrt{2}\,r_C$ is due to the fact that when $L$ is larger than this value the collapse Gaussian gives smaller contributions (see Sec. IV.A for further discussion).}
    \label{ratecube}
\end{figure}
Figure~\ref{ratecube} displays these rates as a function of $L$ for two values of displacement: $\Delta=10^{-3}\,r_C$ (left panel) and $\Delta=10^{3}\,r_C$ (right panel). We first observe that the rate grows with $L$: the larger the object, the more mass comes into play, the faster the collapse. $\Gamma_\text{\tiny{C}}$ (red line) displays a change of slope around $L\simeq\sqrt{2}\,r_C$: this is due to the fact that when $L>\sqrt{2}\,r_C$ the collapse Gaussian gives smaller contributions and the rate grows slower. A further decrease of slope is displayed in the right panel for $L=\Delta$: this is where the states start overlapping and less mass contributes to the rate. A detailed explanation of these behaviours is given in Sec. IV.A. 

In the left panel one sees that $\Gamma_\text{\tiny{A}}$ (blue line) gives a good approximation of the rate for lower values of $L$, while for larger $L$ it grows faster than $\Gamma_\text{\tiny{C}}$, eventually departing from it. Thus the larger the object the worse the approximation given by $\Gamma_\text{\tiny{A}}$. This is expected because for values of $L$ larger than $\Delta$ the states overlap, while $\Gamma_\text{\tiny{A}}$ does not hold in this regime.  We recall that the definition of $\Gamma_\text{\tiny{GPR}}$ holds only for $\Delta\gg r_C$, thus not for the value of $\Delta$ used in the left panel of Fig.~\ref{ratecube}. We show it anyway because $\Gamma_\text{\tiny{GPR}}$ (green line) grasps the correct asymptotic behavior of $\Gamma_\text{\tiny{C}}$: we will come back on this issue later. 

When $\Delta$ is much larger than $r_C$ (right panel), $\Gamma_\text{\tiny{A}}$ gives a good approximation of $\Gamma_\text{\tiny{C}}$ up to $L\simeq r_C$, and it has an offset of about two orders of magnitude for $r_C\lesssim L\lesssim \Delta$. For larger values of $L$, $\Gamma_\text{\tiny{A}}$ departs more and more from $\Gamma_\text{\tiny{C}}$, but this is no issue because $\Gamma_\text{\tiny{A}}$ is not applicable in this regime. $\Gamma_\text{\tiny{GPR}}$ instead gives a good approximation of the exact rate when the body size is $L\gg r_C$ (the only range where $\Gamma_\text{\tiny{GPR}}$ is defined). 
We remark that, in both panels, the actual distances of $\Gamma_\text{\tiny{GPR}}$ and $\Gamma_\text{\tiny{A}}$ from $\Gamma_\text{\tiny{C}}$ depend on the values of $\Delta$ and of the mass density.

Let us now focus on the reduction rate of a cuboidal body with a square face $L_x=L_y=d$ and length $L_z=L$.
\begin{figure}
 \begin{minipage}[h]{0.4\textwidth}
     \includegraphics[width=\textwidth]{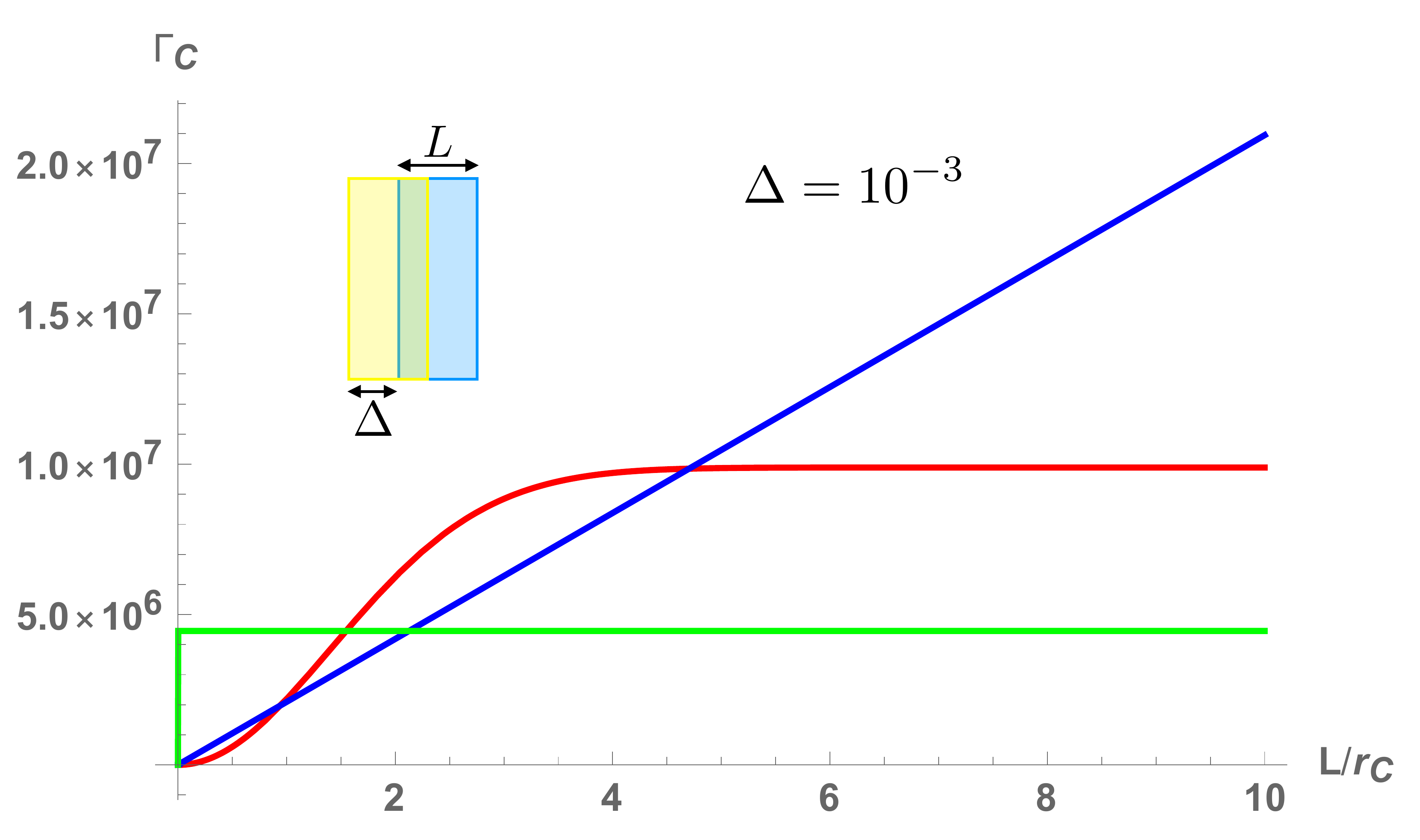} 
  \end{minipage}
  \hspace{1cm}
  \begin{minipage}[h]{0.4\textwidth}
     \includegraphics[width=\textwidth]{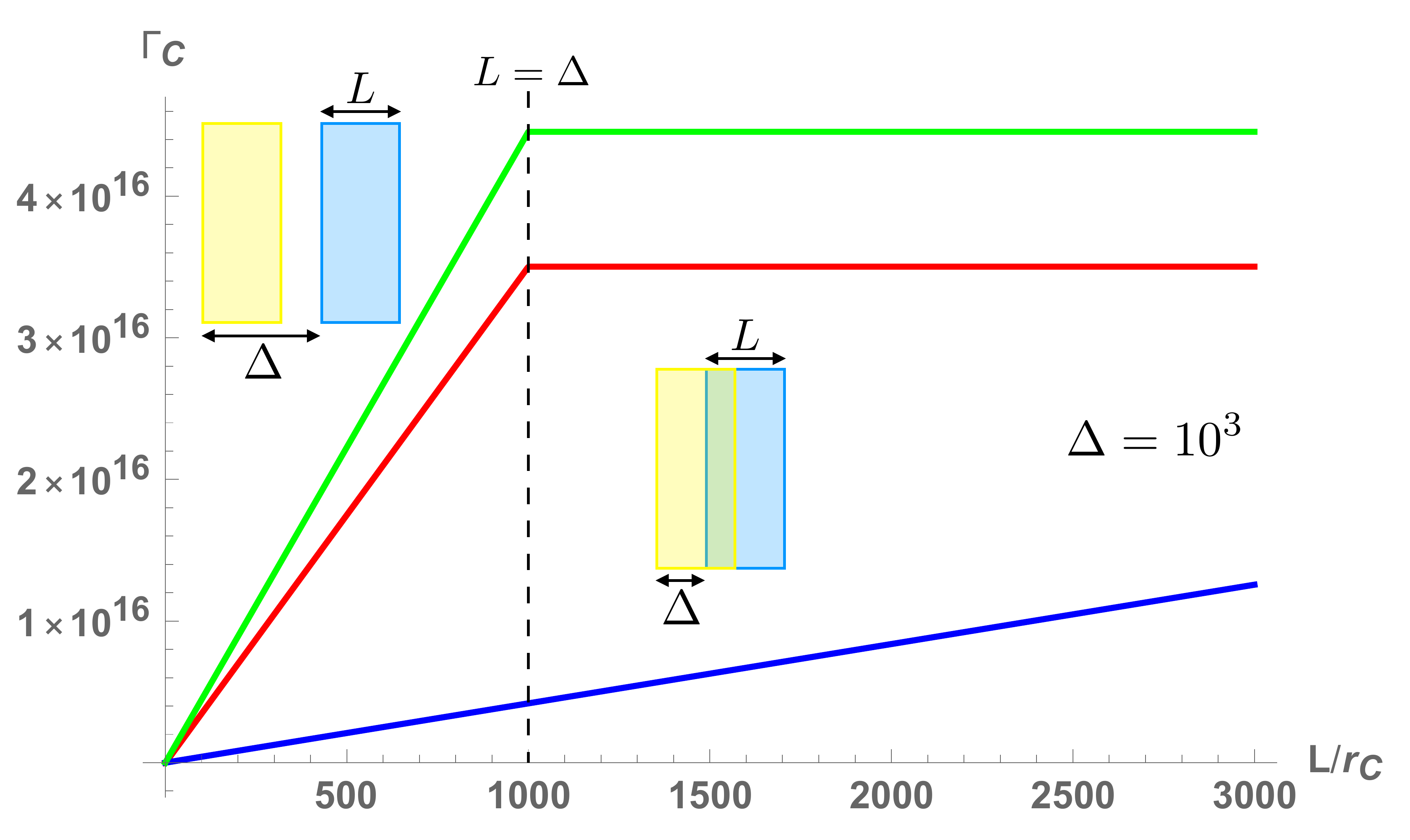}
  \end{minipage}
   \caption{Reduction rate of a cuboidal body as a function of its length $L$ in units of $r_C$, for $\Delta=10^{-3}\, r_C$ (left panel) and $\Delta=10^{3}\,r_C$ (right panel). The lines correspond to: exact reduction rate $\Gamma_\text{\tiny{C}}$ of Eq.~\eqref{gammacuboid} (red line), $\Gamma_\text{\tiny{GPR}}/10^4$ of Eq.~\eqref{gpr} (green line) and $\Gamma_\text{\tiny{A}}$ of Eq.~\eqref{adler} (blue line). Body density is $10^{30}$ nucleons/m$^3$, $r_C=10^{-7}$ m, $\lambda=10^{-8}$ s$^{-1}$, and $d=10\,r_C$. The yellow rectangle represents the state $|q^{\text{\tiny{L}}}\rangle$, the blue rectangle represents  $|q^{\text{\tiny{R}}}\rangle$. In the left panel, the range of $L$ is such that the two states always overlap. In the right panel, the dashed line separates the region where the states do not overlap (left), from the region where they do so (right). A detailed explanation of the behaviour of $\Gamma_\text{\tiny{C}}$ (red line) is given in Fig. 4.}
    \label{ratecuboid}
\end{figure}
Figure~\ref{ratecuboid} compares $\Gamma_\text{\tiny{C}}$, $\Gamma_\text{\tiny{GPR}}$ and $\Gamma_\text{\tiny{A}}$ as a function of the cuboid length $L$, for $\Delta=10^{-3}\,r_C$ (left panel) and $\Delta=10^{3}\,r_C$ (right panel).
We first observe that the exact rate $\Gamma_\text{\tiny{C}}$ (red line) displays a remarkably different behavior than the one for a cubic body (Fig.~\ref{ratecube}): while the reduction rate for the cube grows with the length of its side, the one for the cuboid saturates to a constant value, no matter how long the cuboid is, provided that $L\gtrsim3\sqrt{2}\, r_C$ (see Fig.~\ref{cuboidexplained} for further discussion). As for a cubic body, $\Gamma_\text{\tiny{GPR}}$ (green line) displays the correct asymptotic behavior, also in the regime $\Delta\ll r_C$ where it is not defined. Although $\Gamma_\text{\tiny{A}}$ (blue line) correctly reproduces the exact rate (apart from a numerical factor), it does not have the correct asymptotic behaviour, eventually departing from $\Gamma_\text{\tiny{C}}$ for large values of $L$. This is again explained by the fact that the definition  of $\Gamma_\text{\tiny{A}}$ does not hold for $L>\Delta$.

\subsection{Mass difference effect}
Although the saturation of $\Gamma_\text{\tiny{C}}$ might seem surprising, its origin can be understood by investigating the mathematical properties of the collapse rate. A first important remark is that the collapse rate depends on the \emph{difference of the mass distributions} of the two states onto which the rate is evaluated, as clearly displayed by the definiton~\eqref{rate}. This is a direct consequence of the double commutator in the CSL master equation~\eqref{ME}. The mass difference effect was first discussed, with a different terminology, by Diosi in~\cite{Dio19}.

The mass difference effect is taken into account by the definition of $\Gamma_\text{\tiny{GPR}}$ in Eq.~\eqref{gpr}, which indeed displays the correct asymptotic behavior both in Figs.~\ref{ratecube} and~\ref{ratecuboid} (green line), also in the regime $\Delta\ll r_C$ where it is not defined. On the other side, $\Gamma_\text{\tiny{A}}$ is defined only where the states do not overlap, i.e. it considers only the \emph{total mass} of the body: this explains why for $L>\Delta$ (where the states overlap) $\Gamma_\text{\tiny{A}}$ keeps growing linearly with $L$ (blue line), thus diverging from $\Gamma_\text{\tiny{C}}$.

In order to show that the dependence on the mass difference is responsible for the saturation of the rate displayed in Fig.~\ref{ratecuboid}, we consider Eq.~\eqref{gammacuboid} and we isolate the integrals in the direction of displacement, i.e. the term $g(L_z)-g_{\text{\tiny{$\Delta$}}}(L_z)$. By observing that the function $g(x)$ defined in Eq.~\eqref{gfac} is even, one can easily check that
\beq\label{gg}
g(L_z)-g_{\text{\tiny{$\Delta$}}}(L_z)=g(\Delta)-g_{\text{\tiny{$L_z$}}}(\Delta)\,.
\eeq
The important consequence of this relation is that we can have two equivalent interpretations of the reduction rate in Eq.~\eqref{gammacuboid}, which can be understood either as given by the superposition of two bodies of length $L_z$ at distance $\Delta$, or equivalently as given by the superposition of two bodies of length $\Delta$ at distance $L_z$. 
Interestingly, a relation equivalent to Eq.~\eqref{gg} can be proven for a discrete mass distribution, i.e. for the series in the z direction of $\Gamma_\text{\tiny{D}}$ in Eq.~\eqref{gammaDser}.
\begin{figure}
 \begin{minipage}[h]{0.45\textwidth}
     \includegraphics[width=\textwidth]{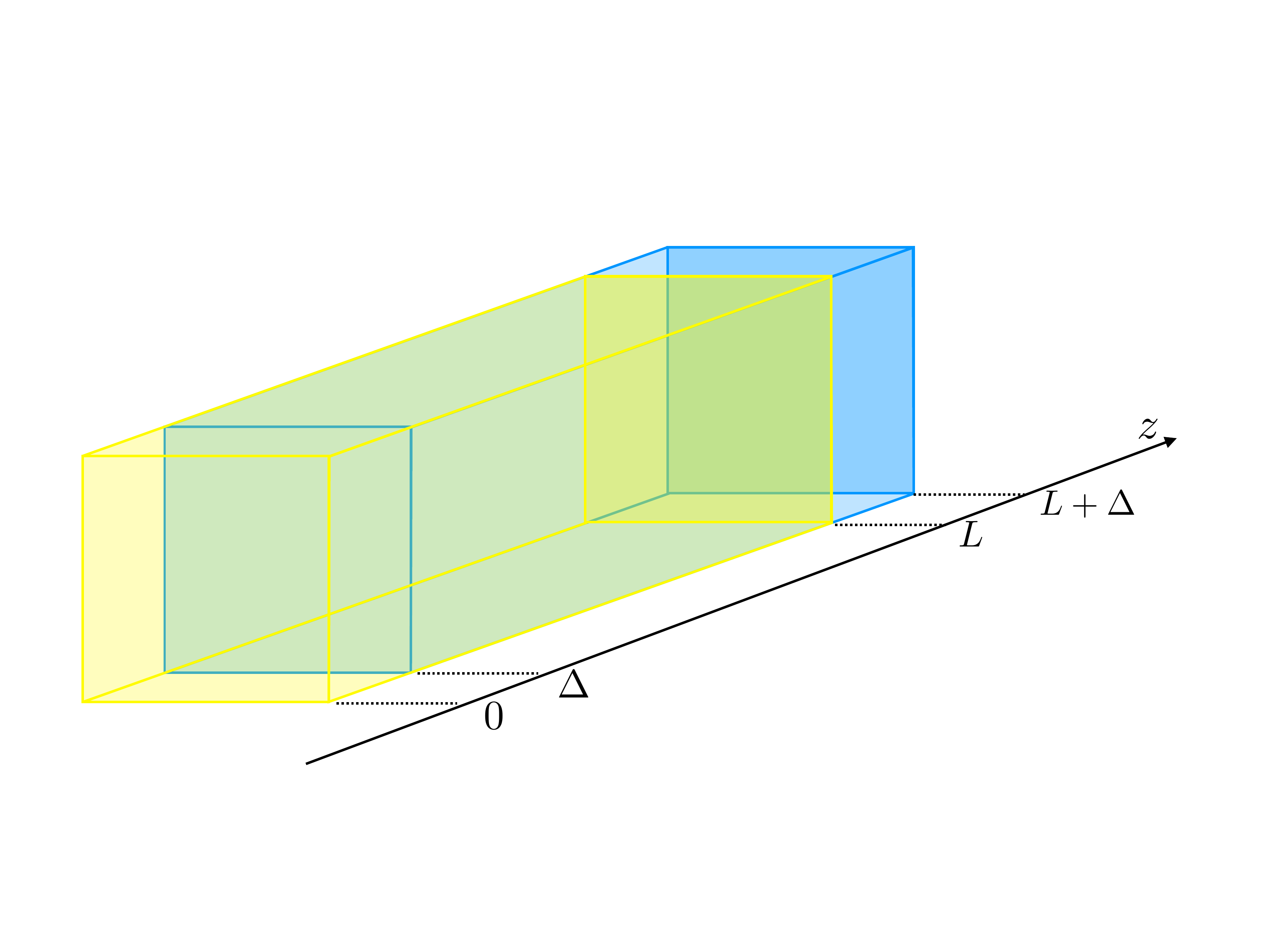} 
  \end{minipage}
  \hspace{1.cm}
  \begin{minipage}[h]{0.45\textwidth}
     \includegraphics[width=\textwidth]{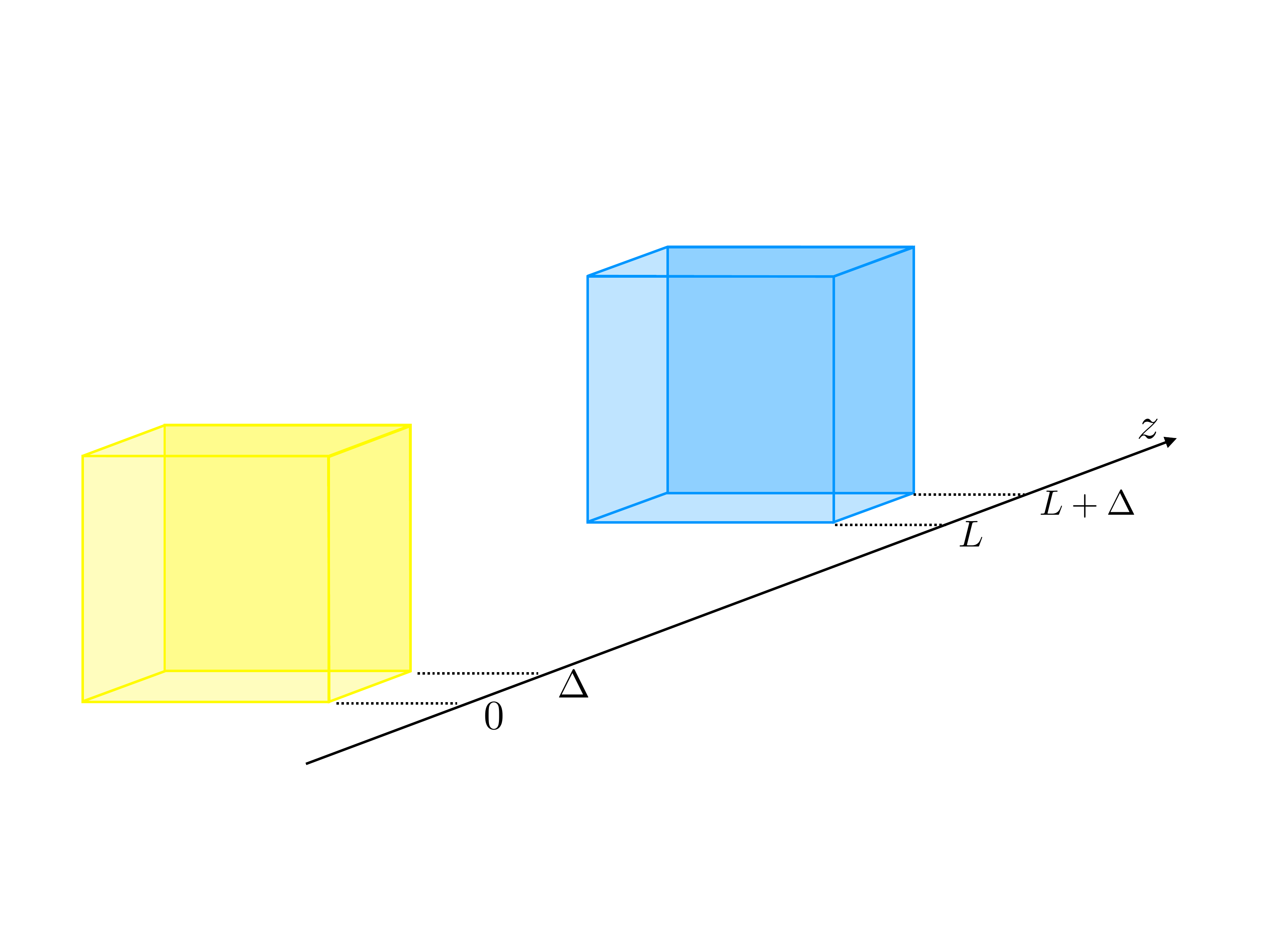}
  \end{minipage}
   \caption{Physical explanation of the mass difference behaviour. The yellow cuboid represents the state $|q^{\text{\tiny{L}}}\rangle$, the blue cuboid represents  $|q^{\text{\tiny{R}}}\rangle$. In the region where the two states overlap (left figure) the mass difference is zero. The right figure shows the mass regions that actually contribute towards the collapse rate, which correspond to a cuboid of length $\Delta$ with superposition distance $L$.}
    \label{massdiff}
\end{figure}
Figure~\ref{massdiff} gives a physical explanation of Eq.~\eqref{gg}: when $\Delta< L_z$, in the region where the two states overlap the mass difference is zero, thus the effective contribution to the rate is the same as that of two bodies of width $\Delta$ at distance $L_z$. As a consequence, in the region $L_z\gg r_C$, even if $L_z$ grows this contribution to the collapse rate stays constant because it depends only on the slice of width $\Delta$. This fact is manifest in the cuboidal collapse rate of Fig.~\ref{ratecuboid}, because in Eq.~\eqref{gammacuboid} the term $g(L_z)-g_{\text{\tiny{$\Delta$}}}(L_z)$ of $\Gamma_\text{\tiny{C}}$ is multiplied by a constant factor $g(d)^2$. Conversely, in the reduction rate for a cubic body this behavior is hidden by the fact that the term $g(L)^2$ in $\Gamma_\text{\tiny{C}}$ grows with $L$.

We are now ready to give a physical interpretation of $\Gamma_\text{\tiny{C}}$ by analysing the red line in the left panel of Fig.~\ref{ratecuboid} in the light of Eq.~\eqref{gammaC}. 
\begin{figure}
 \begin{minipage}[h]{0.5\textwidth}
     \includegraphics[width=\textwidth]{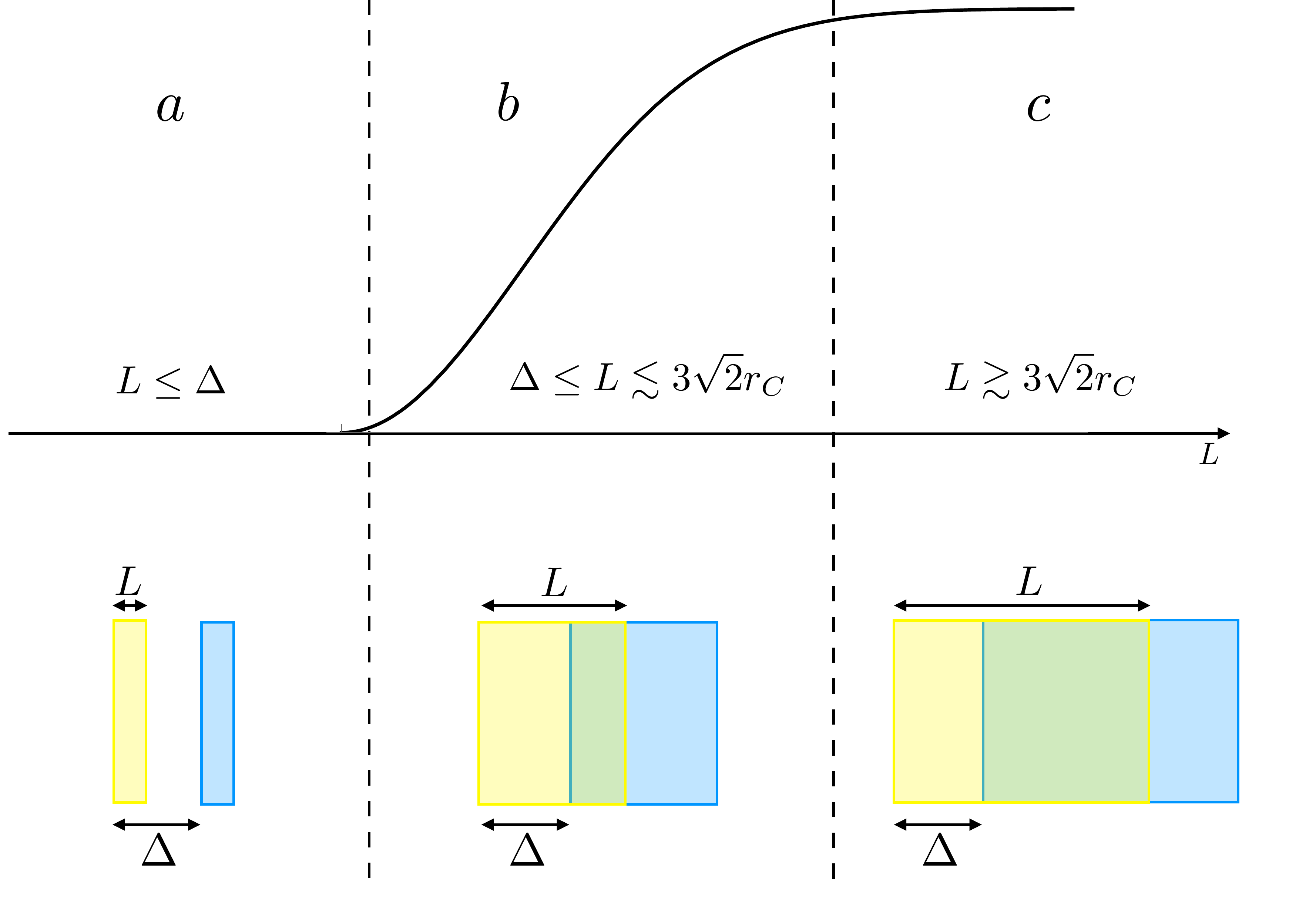} 
  \end{minipage}
   \caption{Physical explanation of the cuboid reduction rate. Top: $\Gamma_\text{\tiny{C}}$ for a cuboidal body as a function of its length in units of $r_C$ (red line in the left panel of Fig.~\ref{ratecuboid}). Bottom: states contributing to the collapse rate in the respective regimes. The yellow rectangle corresponds to $|q_\text{\tiny{L}}\rangle$, the blue rectangle to $|q_\text{\tiny{R}}\rangle$: the region where the two states overlap does not contribute because the mass difference is zero (see Fig.3). 
   }
    \label{cuboidexplained}
\end{figure}
We distinguish three regimes (Fig.~\ref{cuboidexplained}): a) $L\leq\Delta\ll r_C$: the cuboid length $L$ is smaller than the superposition distance $\Delta$, thus the mass difference effect does not take place. 
Since the distance $\Delta$ is much smaller than the variance of the collapse gaussian ($\sqrt{2}\,r_C$), the Gaussian is essentially flat in this region and the rate grows quadratically with $L$. Nonetheless, $L$ is very small in this region, only a tiny amount of mass is involved in the collapse process and the reduction rate is very small.
b) $\Delta\leq L\lesssim 3\sqrt{2} r_C$: the cuboid length $L$ is larger than the superposition distance $\Delta$: the mass difference effect takes place, thus effectively the cuboid length is fixed ($\Delta$) and the superposition distance changes ($L$). The rate grows because  the collapse Gaussian correlates the two terms of the superposition. This happens as long as $L$ is smaller than (about) 3 standard deviations ($3\sqrt{2} r_C$). The larger $L$, the smaller the contribution of the collapse Gaussian, which explains the flattening of the curve. 
c) $L\gtrsim3\sqrt{2} r_C$: also in this region the mass difference effect takes place: the cuboid length is $\Delta$, and the superposition distance is $L$. This is larger than the collapse Gaussian, which in this region is essentially zero. Accordingly, no further contribution is added to the rate, which thus stays constant.

\begin{figure}
 \begin{minipage}[h]{0.5\textwidth}
     \includegraphics[width=\textwidth]{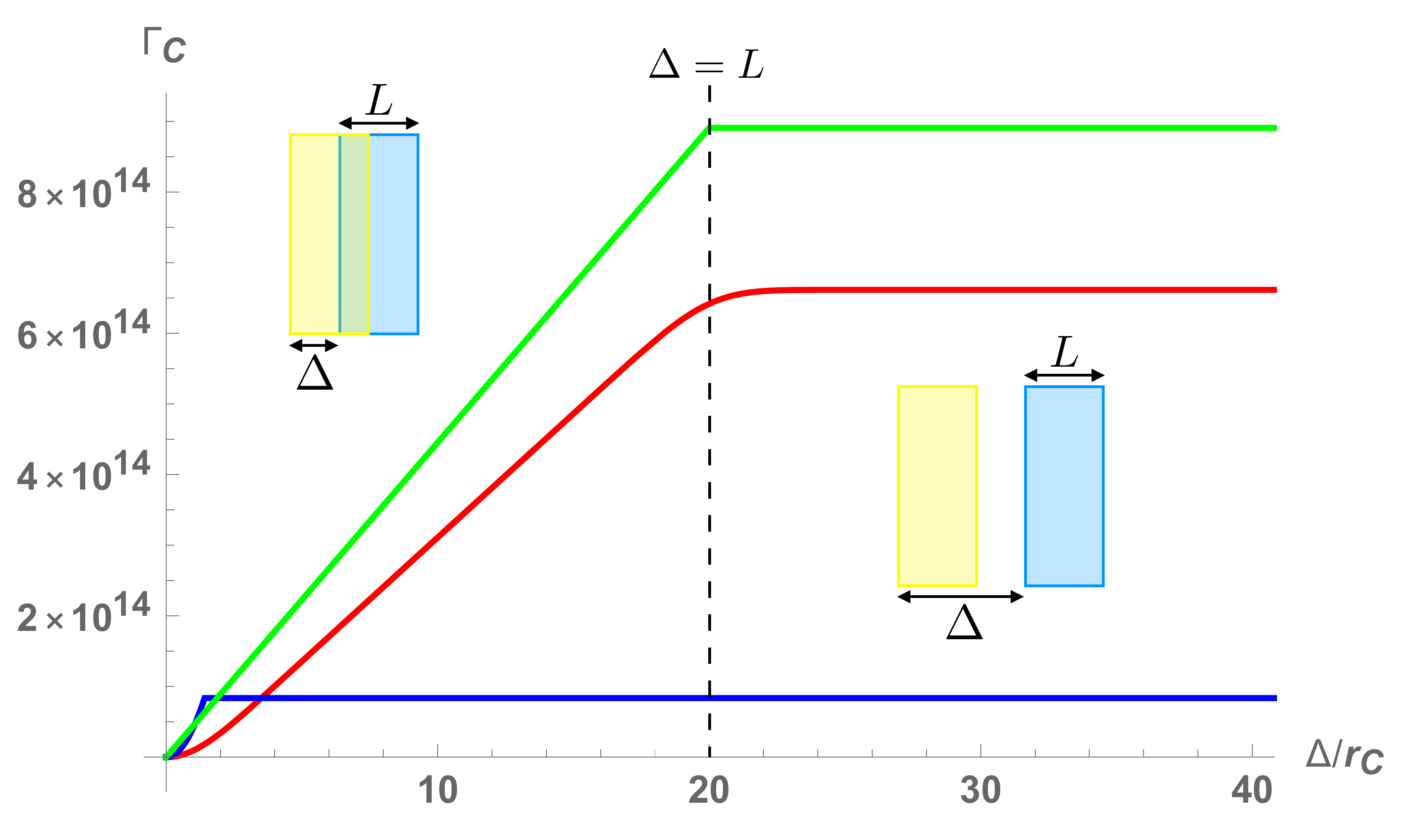} 
  \end{minipage}
   \caption{Reduction rate of a cuboidal body of fixed length $L=20\,r_C$  as a function of the displacement $\Delta$ in units of $r_C$. The lines correspond to: exact reduction rate $\Gamma_\text{\tiny{C}}$ of Eq.~\eqref{gammacuboid} (red line), $\Gamma_\text{\tiny{GPR}}$ of Eq.~\eqref{gpr} (green line) and $\Gamma_\text{\tiny{A}}$ of Eq.~\eqref{adler} (blue line). Body density is $10^{30}$ nucleons/m$^3$, $r_C=10^{-7}$ m, $\lambda=10^{-8}$ s$^{-1}$, $d=10\,r_C$. The yellow rectangle represents the state $|q^{\text{\tiny{L}}}\rangle$, the blue rectangle represents  $|q^{\text{\tiny{R}}}\rangle$. The dashed line separates the region where the states do overlap ($\Delta\leq L$), from the region where they do not ($\Delta>L$).}
    \label{Deltavar}
\end{figure}
Another consequence of the mass difference effect is displayed in Figure~\ref{Deltavar}, which shows the collapse rate for a cuboid of fixed dimension ($L=20\,r_C$, $d=10\,r_C$) as a function of the displacement $\Delta$. We first observe that $\Gamma_\text{\tiny{C}}$ displays the same behaviour as in Fig.~\ref{ratecuboid} and Fig.~\ref{cuboidexplained}: this is explained by the symmetry~\eqref{gg}, according to which the roles of $\Delta$ and $L$ can be interchanged. The reduction rate grows with $\Delta$ because the region where the states overlap decreases and more mass contributes to the rate. When the displacement exceeds the body length ($\Delta>L$) the rate stays constant because it is proportional to the total mass of the body (no overlap between the states). Both $\Gamma_\text{\tiny{GPR}}$ (defined for $\Delta/r_C\gg1$) and $\Gamma_\text{\tiny{A}}$ (defined for $\Delta>L$) well agree with $\Gamma_\text{\tiny{C}}$ (besides numerical factor).

We recall that when the lattice distance is larger than $\sqrt{2} r_C$, the continuous mass approximation is not valid and one needs to use $\Gamma_\text{\tiny{D}}$ (see Sec. III and Appendix B). We stress that in this case the mass difference effect takes places only under special conditions.
\begin{figure}
 \begin{minipage}[h]{0.5\textwidth}
     \includegraphics[width=\textwidth]{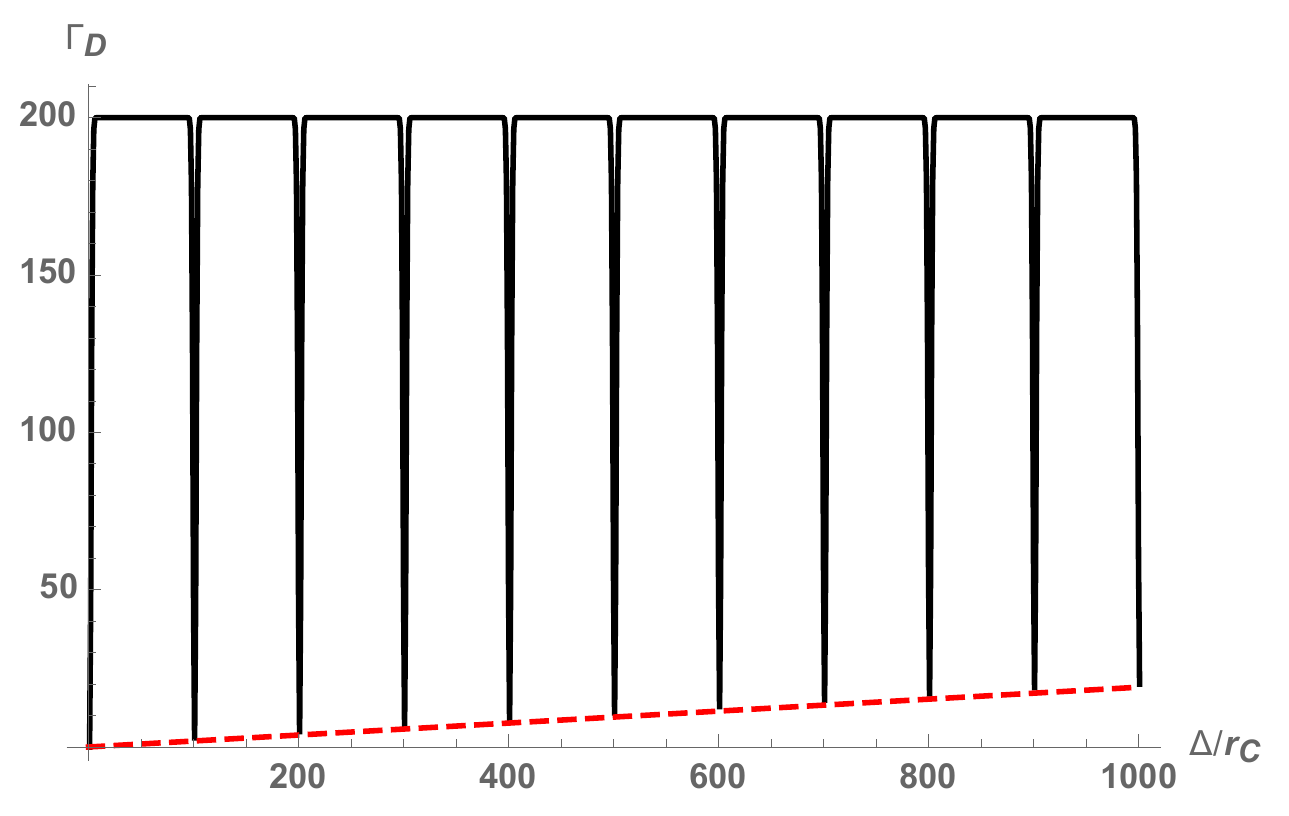} 
  \end{minipage}
   \caption{Reduction rate $\Gamma_\text{\tiny{D}}$ (black solid line) as a function of the displacement $\Delta$ in units of $r_C$. The cuboid is a body of side $d=10^2\,r_C$, length $L=10^3\,r_C$, and lattice constant $l=10^2\,r_C$. The red dashed line passes through the local minima of $\Gamma_\text{\tiny{D}}$. Body density is $10^{15}$ nucleons/m$^3$, $r_C=10^{-7}$ m, $\lambda=10^{-8}$ s$^{-1}$.}
    \label{discrete}
\end{figure}
Figure~\ref{discrete} displays the collapse rate for a cuboidal body with lattice constant $l=10^2\,r_C$,  side $d=10^2\,r_C$ and length $L=10^3\,r_C$ as a function of the displacement $\Delta$. 
The range of $\Delta$ is chosen in such a way that $\Delta\leq L$; in this regime for a continuous mass density one has the physical picture of Fig.~\ref{massdiff}: the two states overlap, the mass difference effect takes place and $\Gamma_\text{\tiny{C}}$ grows with $\Delta$. Figure~\ref{discrete} clearly shows that in the discrete case this does not happen, and $\Gamma_\text{\tiny{D}}$ (black solid line) essentially stays constant. This happens because in general the sites of the state $|q_\text{\tiny{L}}\rangle$ do not overlap with those of the state $|q_\text{\tiny{R}}\rangle$: the mass difference is always non-zero, all the sites contribute to the collapse rate which thus stays constant. However, interestingly the reduction rate experiences sudden drops when the displacement $\Delta$ is an integer multiple of the lattice constant. This is where the mass difference effect takes place for a discrete mass distribution and where one can have a physical picture similar the one depicted in Fig.~\ref{massdiff}: the sites of the state $|q_\text{\tiny{L}}\rangle$ that exactly overlap those of the state $|q_\text{\tiny{R}}\rangle$ do not contribute to the rate because the mass difference is zero. Accordingly, only the sites in the regions that do not overlap contribute, and the reduction rate is proportional to the volume of such regions. This is confirmed by the fact that the local minima grow linearly with $\Delta$, as shown by the red dashed line in Fig.~\ref{discrete}.

\subsection{Other geometries}

As we have mentioned in the previous section, the cuboidal geometry offers many advantages both for the mathematical analysis of the collapse rate, and for its physical understanding. Nonetheless, the results obtained in this section hold also for other simple geometries, like spheres and cylinders. In order to show this, we focus on the experimentally relevant regime $\Delta\ll r_C$, in which case the collapse rate~\eqref{gammacuboid} for a cuboid reduces to
\beq\label{cuboidappr}
\Gamma_\text{\tiny{C}}=\lambda \,N_{\text{\tiny{TOT}}}^2\, g(L_x)g(L_y)\, \frac{\Delta^2}{L_z^2}\left(1-e^{-\frac{L_z^2}{4r_C^2}}\right)\,.
\eeq
In this same regime, the collapse rates for a cylindric body and for a sphere respectively read~\cite{NimHorHam14}
\beqa
\Gamma_\text{\tiny{cyl}}&=&\lambda \,N_{\text{\tiny{TOT}}}^2\, \frac{4 r_C^2}{R_\text{\tiny{cyl}}^2}\left[1-e^{-\frac{R_\text{\tiny{cyl}}^2}{2r_C^2}}\left(I_0\left[\frac{R_\text{\tiny{cyl}}^2}{2r_C^2}\right]+I_1\left[\frac{R_\text{\tiny{cyl}}^2}{2r_C^2}\right]\right)\right]\, \frac{\Delta^2}{L_z^2}\left(1-e^{-\frac{L_z^2}{4r_C^2}}\right)\\
\Gamma_\text{\tiny{sph}}&=&\lambda \,N_{\text{\tiny{TOT}}}^2\, \frac{3 r_C^4}{R_\text{\tiny{sph}}^6}\left[e^{-\frac{R_\text{\tiny{sph}}^2}{r_C^2}}-1+\frac{R_\text{\tiny{sph}}^2}{2r_C^2}\left(e^{-\frac{R_\text{\tiny{sph}}^2}{r_C^2}}+1\right)\right]\Delta^2\,.
\eeqa
\begin{figure}
 \begin{minipage}[h]{0.4\textwidth}
     \includegraphics[width=\textwidth]{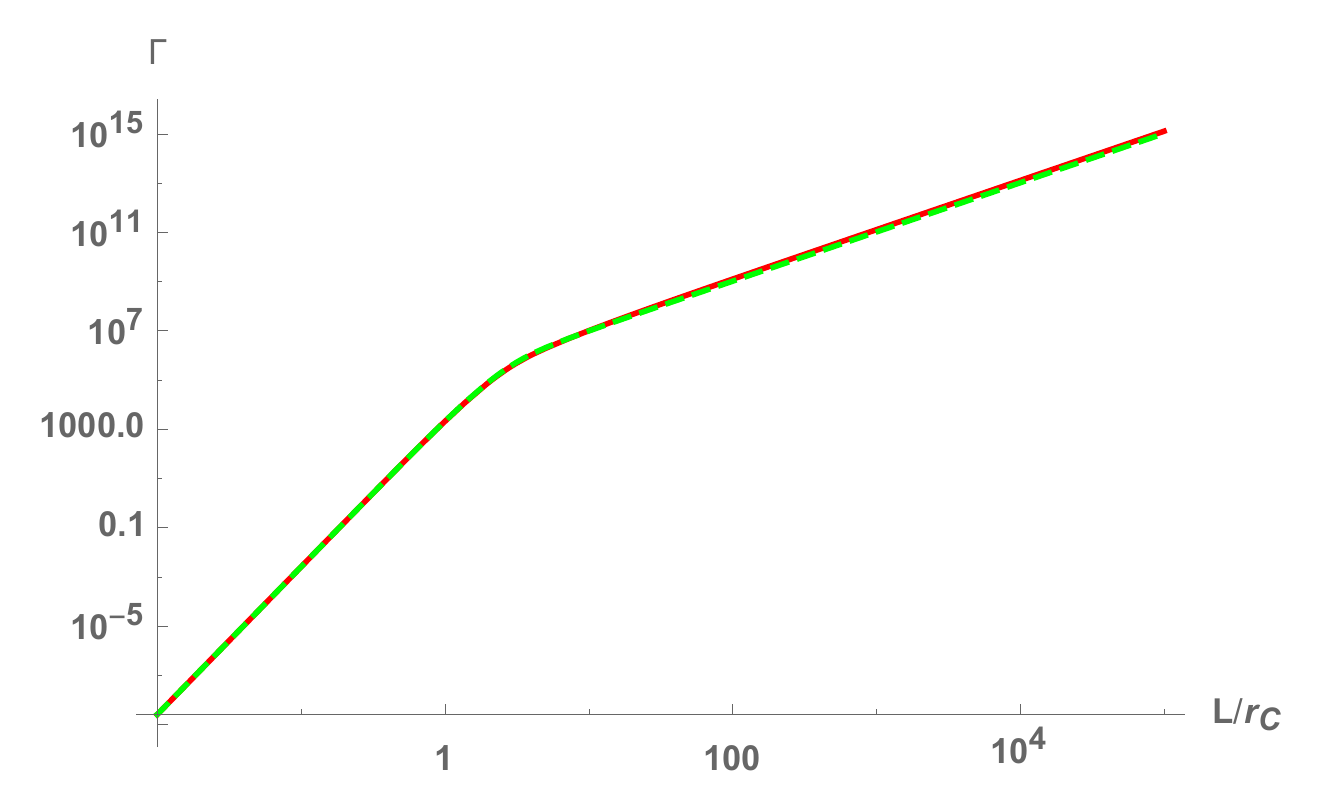} 
  \end{minipage}
  \hspace{1cm}
  \begin{minipage}[h]{0.4\textwidth}
     \includegraphics[width=\textwidth]{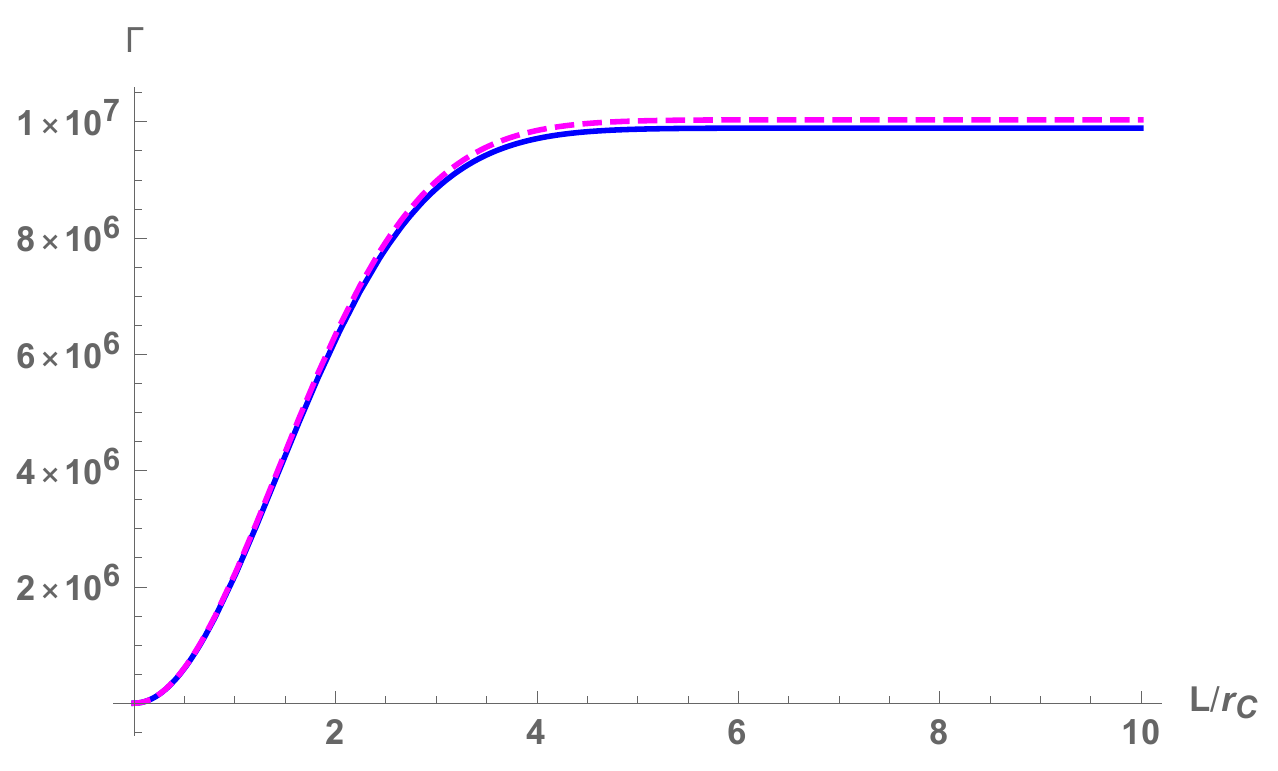}
  \end{minipage}
   \caption{Comparison among collapse rates obtained with different geometries that have the same volume. Left panel: Log-Log plot of the collapse rate for a cube (red solid line) vs sphere (green dashed line). Right panel: linear plot of the rate for a cuboid (blue solid line) vs cylinder (magenta dashed line). Body density is $10^{30}$ nucleons/m$^3$, $r_C=10^{-7}$ m, $\lambda=10^{-8}$ s$^{-1}$, $d=10\,r_C$, $\Delta=10^{-3}\,r_C$. }
    \label{geom}
\end{figure}
Figure~\ref{geom} compares the reduction rates for a cuboid with square face of side $d$ and length $L$ with that for a cylinder with radius $R_\text{\tiny{cyl}}=d/\sqrt{\pi}$ and length $L$ (left panel); and the rate for a cube of side $L$ with the one for a sphere or radius $R_\text{\tiny{sph}}=L \left(3/4\pi\right)^{1/3}$ (right panel). The radii of the cylinder and of the sphere are chosen in such a way that their volumes match those of the cuboid and of the cube respectively.
The plots clearly show that the cubic and cuboidal collapse rates very well describe respectively the rates for spheric and cylindric geometries. 
This is helpful in the scenario when the average distance among particles is such that $\Gamma_\text{\tiny{C}}$ is not a good description of $\Gamma_\text{\tiny{D}}$. One thus needs to resort to a numerical evaluation of $\Gamma_\text{\tiny{D}}$, which can be performed quite easily for a cuboidal geometry, but it is rather cumbersome for other geometries.

\section{Layering effect}
In a recent paper~\cite{CarVinBas18} the idea was put forward that a body with a multilayer structure has a larger diffusion coefficients~\eqref{etadisc} than a uniform one, and few case studies were numerically analysed. Later this idea was exploited in a cantilever experiment to improve the bounds on the collapse parameters~\cite{lay-exp}. In this section we investigate such a \emph{layering effect} and we show that this is a consequence of the \emph{mass difference effect} previously discussed. Both for simplicity and to allow for a comparison with the above mentioned papers, we consider a cuboidal mass distribution with sides $L_x$, $L_y$, $L_z$ and average density $\varrho$.
Let us first consider the master equation~\eqref{MEcmsD} and let us focus on the diffusion coefficient along the z direction defined in Eq.~\eqref{gammasmall} that we report here in a slightly different version
\beq
\eta^{zz}=\frac{\lambda}{2\,m_N^2}\int d^3u\int d^3v\,\mu(\boldsymbol{u}) \mu(\boldsymbol{v}) \frac{\partial}{\partial u_z}\frac{\partial}{\partial v_z} \left(e^{-\frac{(\boldsymbol{u}-\boldsymbol{v})^2}{4r_C^2}}\right)\,.
\eeq
In order to understand the physical meaning of this diffusion coefficient, we integrate by parts along the z direction obtaining
\beq\label{etazzvar}
\eta^{zz}=\frac{\lambda}{2\,m_N^2}\int d^3u\int d^3v\, \nu(\boldsymbol{u})\nu(\boldsymbol{v}) e^{-\frac{(\boldsymbol{u}-\boldsymbol{v})^2}{4r_C^2}}\,,
\eeq
where
\beq\label{nu}
\nu(\boldsymbol{u})=2 [\delta(u_z)-\delta(u_z-L_z)]\mu(\boldsymbol{u})+\frac{\partial \mu(\boldsymbol{u})}{\partial u_z}
\eeq
We then see that $\eta^{zz}$ measures the \emph{correlation of the variation of the mass density} along the z direction, averaged by a Gaussian distribution of width $\sqrt{2}\,r_C$. The first term of Eq.~\eqref{nu} is a boundary contribution, that measures the mass variation at the body's boundaries. 
The last term depends on the  variation of the mass density along the body: when this is uniform, this term does not contribute; when the internal distribution is not uniform, like in the multilayered case, this term gives further contributions, thus increasing the diffusion coefficient.

\begin{figure}
 \begin{minipage}[h]{0.45\textwidth}
     \includegraphics[width=\textwidth]{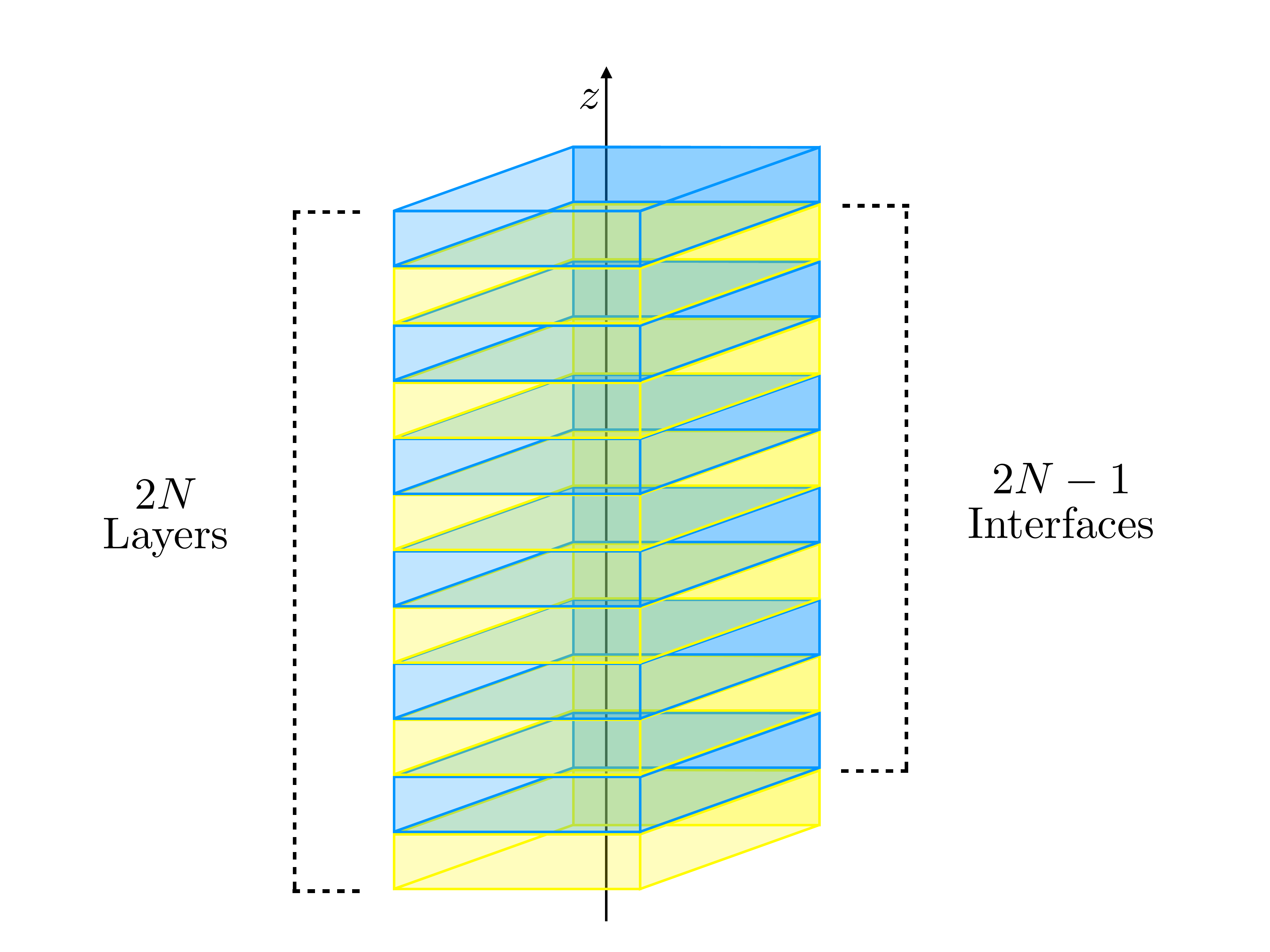} 
  \end{minipage}
  \hspace{1.cm}
  \begin{minipage}[h]{0.45\textwidth}
     \includegraphics[width=\textwidth]{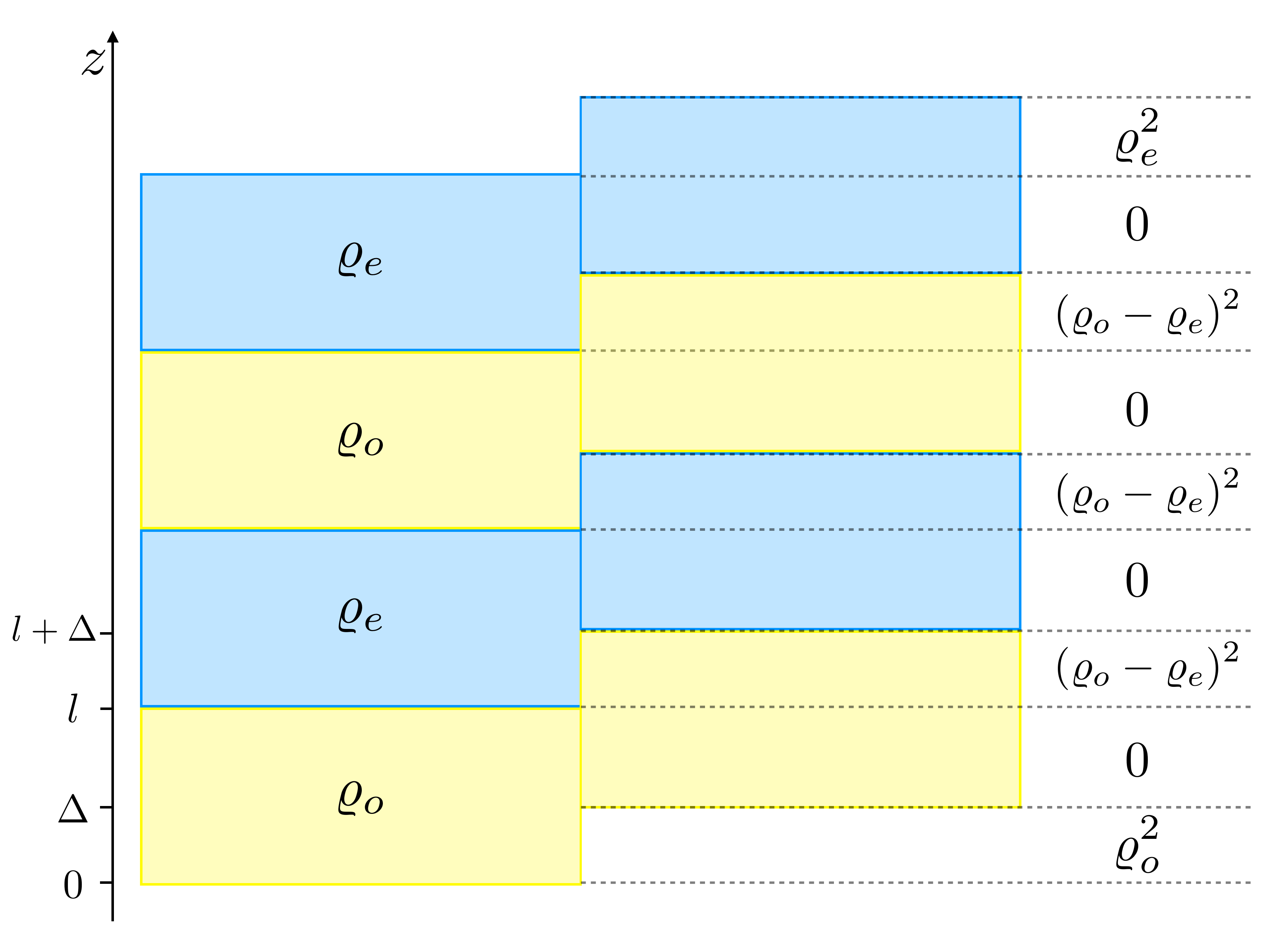}
  \end{minipage}
   \caption{Left panel: special case of a layered object with $2N$ layers all having the same thickness, and two different mass densities: $\varrho_o$ (yellow layer) and $\varrho_e$ (blue layer). The layering effect is proportional to the number of interfaces ($2N-1$), each of which contributes with the square of the mass difference between two consecutive layers (right panel).}
    \label{layered}
\end{figure}

We stress once more that this is a property that belongs to the master equation~\eqref{MEcmsD}, thus affecting both the fluctuations of the dynamics (measured e.g. by the density noise spectrum, like in cantilever experiments~\cite{lay-exp}), and 
the reduction rate, which summarizes the evolution of the off-diagonal elements of the density matrix. As we showed in Sec. II (see Eq.~\eqref{gammasmall}), when the distance between the off-diagonal elements is $\Delta\ll r_C$, the collapse rate is related to the diffusion coefficient by the formula $\Gamma_\text{\tiny{C}}=\eta^{zz}\Delta^2$ . 
In this regime the collapse rate thus measures the correlation of the mass difference of the two states onto which it is evaluated. The right picture of Fig.~\ref{layered} gives a physical intuition of the terms contributing to the reduction rate for a layered object: besides the boundary contributions (proportional to $\varrho_o^2$ and $\varrho_e^2$), there are additional terms that are proportional to the mass difference among the layers. These terms are not present for a uniform body since the mass difference inside the body is zero.  The former analysis gives physical context to the elegant proof in terms surface tensors given by Diosi in~\cite{Dio19}, who showed that the diffusion factor $\eta^{zz}$ encodes a \emph{surface effect}. In the reminder of this section we further give a quantitative estimate of this effect.

We are now ready to consider a layered cuboidal body with square faces of side $d$ (x-y plane) and length $L$ (left panel of Fig.~\ref{layered}). 
The body has $2N$ layers in the z direction, all having a different uniform density $\varrho_i$ and different thickness $l_i$, in such a way that $\sum_{i=1}^{2N}l_i=L$. We start from the definition in Eq.~\eqref{gammasmall}, which we factorize in the three directions:
\beq\label{etazzfact}
\eta^{zz}=\frac{\lambda\,d^4}{4r_C^2\,m_N^2}g(d)^2\int du_z\int dv_z\,\mu(u_z) \mu(v_z) \left[1-\frac{(u_z-v_z)^2}{2r_C^2}\right] e^{-\frac{(u_z-v_z)^2}{4r_C^2}}\,,
\eeq
where $g(x)$ is defined in Eq.~\eqref{gfac},
\beq
\mu(u_z)=\sum_{i=1}^{2N}\varrho_i \,\Theta(u_z-l_{i-1})\,\Theta(l_i-u_z)\,,
\eeq
$l_0=0$, and $\Theta(x)=1$ if $x\geq0$, and zero elsewhere. The double integral in Eq.~\eqref{etazzfact} can be performed exactly giving
\beq
\eta^{zz}=\frac{\lambda\,d^4}{2\,m_N^2}g(d)^2\sum_{i,j=1}^{2N}\varrho_i \,\varrho_j \,\left(e^{-\frac{(l_i-l_j)^2}{4r_C^2}}-e^{-\frac{(l_i-l_{j-1})^2}{4r_C^2}}-e^{-\frac{(l_{i-1}-l_j)^2}{4r_C^2}}+e^{-\frac{(l_{i-1}-l_{j-1})^2}{4r_C^2}}\right)\,.
\eeq
In order to simplify the analysis, we consider two types of layers that alternate in the body: layers with odd index have thickness $l_o$ and density $\varrho_o$, layers with even index are respectively characterised by $l_e$ and $\varrho_e$. This allows to rewrite the previous equation as follows:
\beq\label{etazzeo}
\eta^{zz}=\frac{\lambda\,d^4}{2\,m_N^2}g(d)^2\left[\varrho_o^2+\varrho_e^2-2\varrho_o\varrho_ee^{-\frac{L^2}{4r_C^2}}+\Delta\varrho^2\sum_{i,j=0}^{2N-1}(-1)^{i-j}e^{-\frac{(l_i-l_j)^2}{4r_C^2}}-\Delta\varrho^2\right]\,
\eeq
where $\Delta\varrho\equiv(-1)^{i+1}(\varrho_i-\varrho_{i+1})$ is the density difference between two consecutive layers. We thus see that this equation has the same structure as Eqs.~\eqref{etazzvar}-\eqref{nu}: the first term inside the square brackets corresponds to the contribution from the boundary, while the other two terms measure the variation of mass density inside the body. It is evident that when the body has uniform density ($\Delta\varrho=0$) these terms vanish. Another important remark is that the leading order ($i=j$) of the sum in Eq.~\eqref{etazzeo} does not depend on the thickness of the layers.
This comes as no surprise since $\eta^{zz}$ depends the \emph{derivative of the mass density} (i.e. its variation over an infinitesimal distance) at the interface between two layers, thus the layers' thickness play no major role. We specialize to the case where all layers have the same thickness $l$, which allows to write the leading ($|i-j|=0$) and first ($|i-j|=1$) orders of $\eta^{zz}$ respectively as follows
\beqa
\eta^{zz}_{(0)}&=&\frac{\lambda\,d^4}{2\,m_N^2}g(d)^2\Big[(2N-1)\,\Delta\varrho^2+\varrho_o^2+\varrho_e^2\Big]\label{etazz0}\\
\eta^{zz}_{(1)}&=&\frac{\lambda\,d^4}{2\,m_N^2}g(d)^2\left[-2 (2N-1)\,\Delta\varrho^2\,e^{-\frac{l^2}{4r_C^2}}\right]\label{etazz1}\,.
\eeqa
Equation~\eqref{etazz0} shows that the main contribution of having a layered object is given by the difference of mass density $\Delta\varrho$ at the $2N-1$ interfaces between the layers. In order to maximize $\eta^{zz}$ one should choose $\Delta\varrho$ and $N$ to be as large as possible. The first order~\eqref{etazz1} has the tendency to decrease $\eta^{zz}$ (negative sign), therefore the layer thickness should be chosen to be $l\gtrsim r_C$, in order to minimize the first order. In general the larger $l$ the better, but in experimental situations where the total size of the body is limited, one should choose $l$ compatibly with the fact of having as many layers as possible. Higher orders are negligible because they decay faster than $\eta^{zz}_{(1)}$.

In order to estimate when it is convenient to exploit a layered object instead of a uniform one, we consider a layered body whose layer thickness $l$ is such that Eq.~\eqref{etazz1} gives a negligible contribution, and the diffusion coefficient $\eta^{zz}_{lay}$ for such a body is given by Eq.~\eqref{etazz0}. In order to make a fair comparison we consider a uniform object that has the same mass and volume as the layered one, i.e. with uniform mass density $\varrho_{uni}=(\varrho_o+\varrho_e)/2$. 
By replacing this in Eq.~\eqref{etazz0} one finds that
\beq
\frac{\eta^{zz}_{lay}}{\eta^{zz}_{uni}}=1+\frac{\left(4N-1\right)\,\Delta\varrho^2}{(\varrho_o+\varrho_e)^2}\,,
\eeq
according to which $\eta^{zz}_{lay}$ is appreciably larger than $\eta^{zz}_{uni}$ when $N\gtrsim (\varrho_o+\varrho_e)^2/4\,\Delta\varrho^2$.

Since the layering effect depends linearly on the number of layers, the larger the body, the more one can benefit of the layering effect. For example, we analyze the cantilever experiment performed in~\cite{lay-exp} using a test mass with the following features: 24 layers of WO$_3$ ($\varrho_o\simeq7.2\times10^3$~kg/m$^3$) alternated with 23 layers of SiO$_2$ ($\varrho_e\simeq2.2\times10^3$~kg/m$^3$), mean layer thickness $l\simeq3.7 \times 10^{-7}$~m, sides $L_x\simeq1.1\times10^{-4}$~m and $L_y\simeq8.2\times10^-5$~m. 
A uniform test body of same size and mass as the layered one must have an uniform density of about $\varrho_{uni}\simeq4.8\times10^3$~kg/m$^3$. One can then estimate that $\eta^{zz}_{lay}/\eta^{zz}_{uni}\simeq2.8\times10$, i.e. that the layered geometry of the resonator is responsible for about one order of magnitude of the overall improvement on the bound of the collapse parameters obtained in~\cite{lay-exp}. Nonetheless, one can estimate that gravitational waves experiments would benefit of a much larger improvement thanks to the layered geometry. The Advanced LIGO interferometer involves a silica cylinder of length $L=2\times 10^{-1}$ m and density $\varrho=2.2\times10^3$ Kg/m$^3$~\cite{LIGO}. If one considers a layered cylinder with $N=10^5$ layers of thickness $l=2\times 10^{-6}$ m and  $\Delta\varrho=5\times10^3$, one finds that $\eta^{zz}_{lay}\simeq10^5\,\eta^{zz}_{uni}$. The LISA Pathfinder involves a cubic alloy of AuPt ($\varrho\simeq2\times10^3$ Kg/m$^3$) of side $L=4.6\times 10^{-2}$ m~\cite{LISA}. The corresponding layered object would fit $N=2.3\times 10^{4}$ layers, and assuming the same $\Delta\varrho$ one finds that $\eta^{zz}_{lay}\simeq10^4\,\eta^{zz}_{uni}$. We thus see that in both experiments the layered structure would largely improve the sensitivity of these experiments of collapse effects.

\section{Conclusions}
We have investigated the properties of the CSL collapse rate for rigid bodies. By exploiting the Euler-Maclaurin formula, we showed that the rate computed for a continuous mass distribution accurately reproduces the exact rate (i.e. the one for a point-like distribution) whenever the average particle distance $l$ is smaller than the width of the collapse Gaussian ($\sqrt{2}\,r_C$). For standard matter, where $l$ is of the order of $10^{-10}$ m, the continuous description is extremely accurate. 

We then focused on the reduction rate for cuboidal bodies, and we compared the exact rate $\Gamma_\text{\tiny{C}}$ with the estimates proposed by Ghirardi, Pearle, Rimini ($\Gamma_\text{\tiny{GPR}}$)~\cite{CSL}, and by Adler ($\Gamma_\text{\tiny{A}}$)~\cite{Adl07}. We found that, in its range of definition, $\Gamma_\text{\tiny{GPR}}$  well approximates the exact rate (besides numerical factor). Also $\Gamma_\text{\tiny{A}}$ is generally close to $\Gamma_\text{\tiny{C}}$, although for some values of the parameters the two rates can differ of few orders of magnitude (see right panel of Fig.~\ref{ratecube}). We further showed that the behaviour of the reduction rate strongly depends on the \emph{mass difference effect}, namely the fact that rate depends on the mass difference of the two states onto which it is evaluated. This peculiar feature of the collapse rate originates from the fundamental properties of the CSL model~\cite{CSL}: the identity of particles, the collapse operator proportional to the mass density, and the double commutator in the CSL master equations~\eqref{ME} and \eqref{MEcm}. We remark that when we expand the c.o.m. master equation~\eqref{MEcm} for small c.o.m. displacements to obtain Eq.~\eqref{MEcmsD}, the  \emph{mass difference effect} becomes a  \emph{mass variation effect} (i.e. a difference over an infinitesimal distance). This is fully encoded in the diffusion coefficients of Eq.~\eqref{MEcmsD}, as explained in Sec. V. 

To complete our analysis of the collapse rate we showed that for discrete mass distributions the mass difference effect takes place only when the displacement among the states is an integer multiple of the lattice constant. We also showed that our results do not strictly depend on the cuboidal geometry, and hold also for spherical and cylindrical geometries.

We then investigated the collapse rate for a layered object.
We showed that a geometry of this kind benefits of the mass difference effect in a way that is proportional to the number of layers and to the square of the mass density difference between consecutive layers. This is an intrinsic property of the diffusion coefficient $\eta^{zz}$ displayed by the master equation~\eqref{MEcmsD}, which measures the \emph{variation of the mass density} along the direction of layering ($z$). Our analysis gives a more solid ground to the idea put forward by Diosi in~\cite{Dio19} that the diffusion factor is a surface effect.


\section*{Acknowledgements}
The authors acknowledge financial support from the H2020 FET Project TEQ (grant n. 766900). AB also acknowledges financial support from INFN, FQXi, and the COST Action QTSpace (CA15220).

\section*{Appendix A: Reduction rate in the momentum space}

In this Appendix we repeat the analysis of Sec. II and we provide the formulas in the momentum space.
The CSL master equation in the momentum space reads
\beq\label{lindmom}
\frac{d}{dt}\hat{\rho}(t)=-\frac{i}{\hbar}[\Ho,\ro(t)]-\frac{\lambda r_C^3}{2\pi^{3/2}m_N^2}\int d^3k\, e^{-r_C^2\boldsymbol{k}^2}[\hat{\mu}(\boldsymbol{k}),[\hat{\mu}(-\boldsymbol{k}),\ro(t)]]\,,
\eeq
where we have introduced the Fourier transform of the mass density operator:
\begin{equation}\label{mumom}
\hat{\mu}(\boldsymbol{k})\equiv\int d^3x\,e^{-i\boldsymbol{k}\cdot\boldsymbol{x}}\hat{\mu}(\boldsymbol{x})=\sum_im_i\,e^{-i\boldsymbol{k}\cdot\hat{\boldsymbol{q}}_i}\,.
\end{equation}
Equation~\eqref{lindmom} allows to rewrite the collapse rate~\eqref{rate} as follows
\beqa\label{ratemom}
\Gamma(q^{\text{\tiny{L}}},q^{\text{\tiny{R}}})&=&\frac{\lambda r^3_C}{2\pi^{3/2}m_N^2}\int d^3k\, e^{-r_C^2k^2} \Big(\mu^{\text{\tiny{L}}}(\boldsymbol{k})-\mu^{\text{\tiny{R}}}(\boldsymbol{k})\Big)\Big(\mu^{\text{\tiny{L}}}(\boldsymbol{-k})-\mu^{\text{\tiny{R}}}(\boldsymbol{-k})\Big)\,.
\eeqa
We rewrite the particles' position operators in terms of the c.o.m. and relative coordinates ($\bqo_i=\bQo+\bro_i$). Under the assumption of rigid body, according to which the relative coordinates are sharply localised (with respect to $r_C$) around the classical positions $\br_i$, i.e. $\langle(\bro_i-\br_i)^2\rangle\ll r_C$, one finds that the c.o.m. master equation reads
\beq\label{lindmomrel}
\frac{d}{dt}\hat{\rho}_{\text{\tiny{CM}}}(t)=-\frac{i}{\hbar}[\Ho_{\text{\tiny{CM}}},\ro_{\text{\tiny{CM}}}(t)]-\frac{\lambda r_C^3}{2\pi^{3/2}m_N^2}\sum_{i,j=1}^Nm_im_j\int d^3k\, e^{-r_C^2\boldsymbol{k}^2}\,e^{-i\bk\cdot(\br_i-\br_j)}\left[e^{-i\bk\cdot\bQo},\left[e^{i\bk\cdot\bQo},\ro_{\text{\tiny{CM}}}(t)\right]\right]\,,
\eeq
where $\hat{\rho}_{\text{\tiny{CM}}}$ and $\Ho_{\text{\tiny{CM}}}$ denote respectively the density matrix and the Hamiltonian of the c.o.m..
By expanding the exponentials for small $\bQo$ and exploiting the relation $\mu(\boldsymbol{-k})=\mu^*(\boldsymbol{k})$ one finds
\beqa\label{lindmomapp}
\frac{d}{dt}\hat{\rho}_{\text{\tiny{CM}}}(t)=-\frac{i}{\hbar}[\Ho_{\text{\tiny{CM}}},\ro_{\text{\tiny{CM}}}(t)]-\frac{\lambda r_C^3}{2\pi^{3/2}m_N^2}\int d^3k\, e^{-r_C^2\boldsymbol{k}^2}\,|\mu(\boldsymbol{k})|^2\left[\bk\cdot\bQo,\left[\bk\cdot\bQo,\ro_{\text{\tiny{CM}}}(t)\right]\right]\,,
\eeqa
which eventually leads to Eq.~\eqref{MEcmsD} with the diffusion coefficients
\beq\label{etafour}
\eta_{\alpha\beta}=\frac{\lambda r_C^3}{2\pi^{3/2}m_N^2}\int d^3k\, e^{-r_C^2\boldsymbol{k}^2}\,|\mu(\boldsymbol{k})|^2\,k_\alpha k_\beta\,.
\eeq
These formulas are those most often used in the literature on the topic~\cite{NimHorHam14,Caretal16,Vinetal17,CarFerBas18}.

\section*{Appendix B: Euler-Maclaurin formula}
In this appendix we show how the Euler-Maclaurin (EM) formula~\eqref{EM} allows to estimate the error made when one estimates the reduction rate with $\Gamma_{\text{\tiny{C}}}$ instead of $\Gamma_{\text{\tiny{D}}}$. We start with a more mathematical definition of the EM formula~\cite{apostol}:

\noindent {\bf Theorem.} \emph{For any function $f(x)$ with a continuous derivative of order $2p+1$ on the interval $[0,N]$, the following identity holds:
\beq\label{EMappB}
\sum_{i=1}^Nf(i)=\int_0^Ndx f(x)+\frac{1}{2}\left[f(N)-f(0)\right]+\sum_{k=1}^{p}\frac{B_{2k}}{2k!}\left[f^{(2k-1)}(N)-f^{(2k-1)}(0)\right]+R_p\,,
\eeq
where $B_k$ is the $k$-th Bernoulli number, and $f^{(k)}$ is the $k$-th derivative of $f(x)$. The reminder $R_p$ is defined as follows
\beq
R_p\equiv\frac{1}{(2p+1)!}\int_0^Ndx\,P_{2p+1}(x)f^{(2p+1)}(x)\,,
\eeq
where $P_k(x)$ is the periodic Bernoulli function of the $k$-th order.} 

Equation~\eqref{EMappB} holds for any integer $p\geq0$, which sets the order of approximation of the error estimate, and can be chosen in such a way to minimize the remainder $R_p$. 
However, we remark that choosing a larger $p$ does not necessarily correspond to a better error estimate, i.e. to a smaller $R_p$.

\begin{figure}
 \begin{minipage}[h]{0.45\textwidth}
     \includegraphics[width=\textwidth]{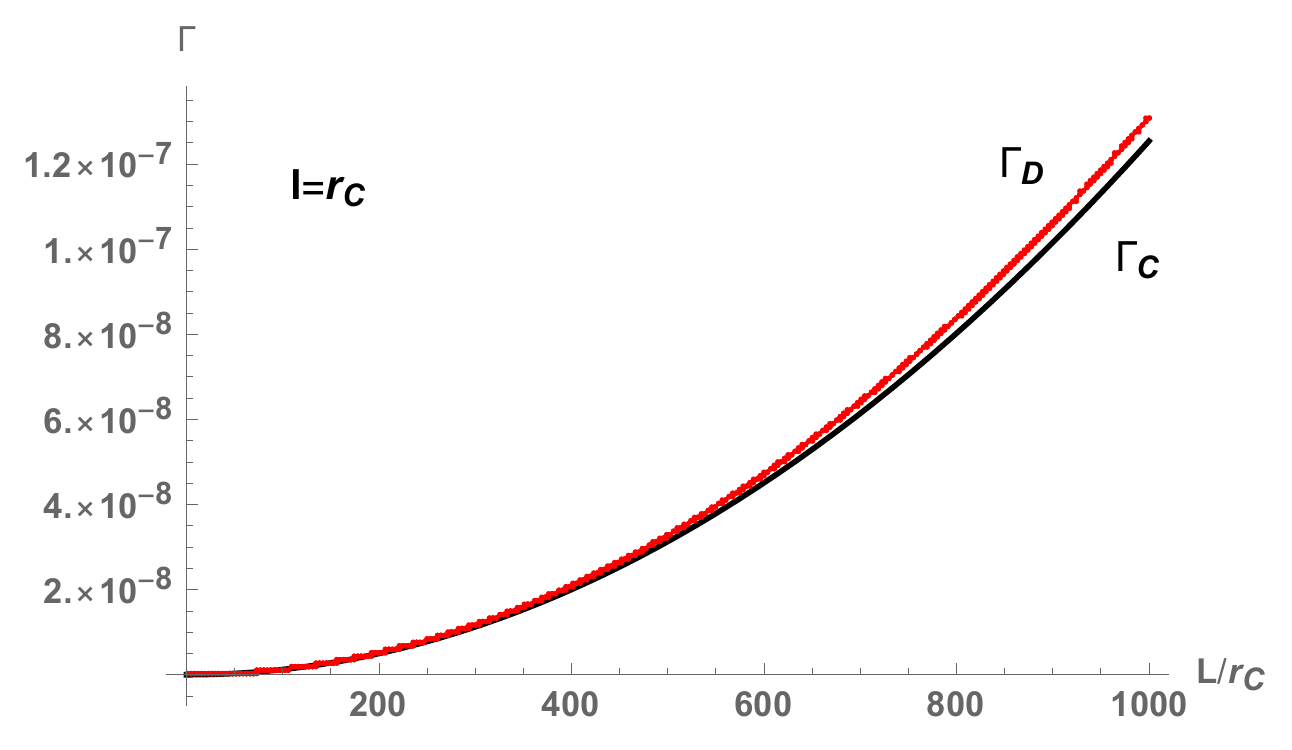} 
  \end{minipage}
  \hspace{1.cm}
  \begin{minipage}[h]{0.45\textwidth}
     \includegraphics[width=\textwidth]{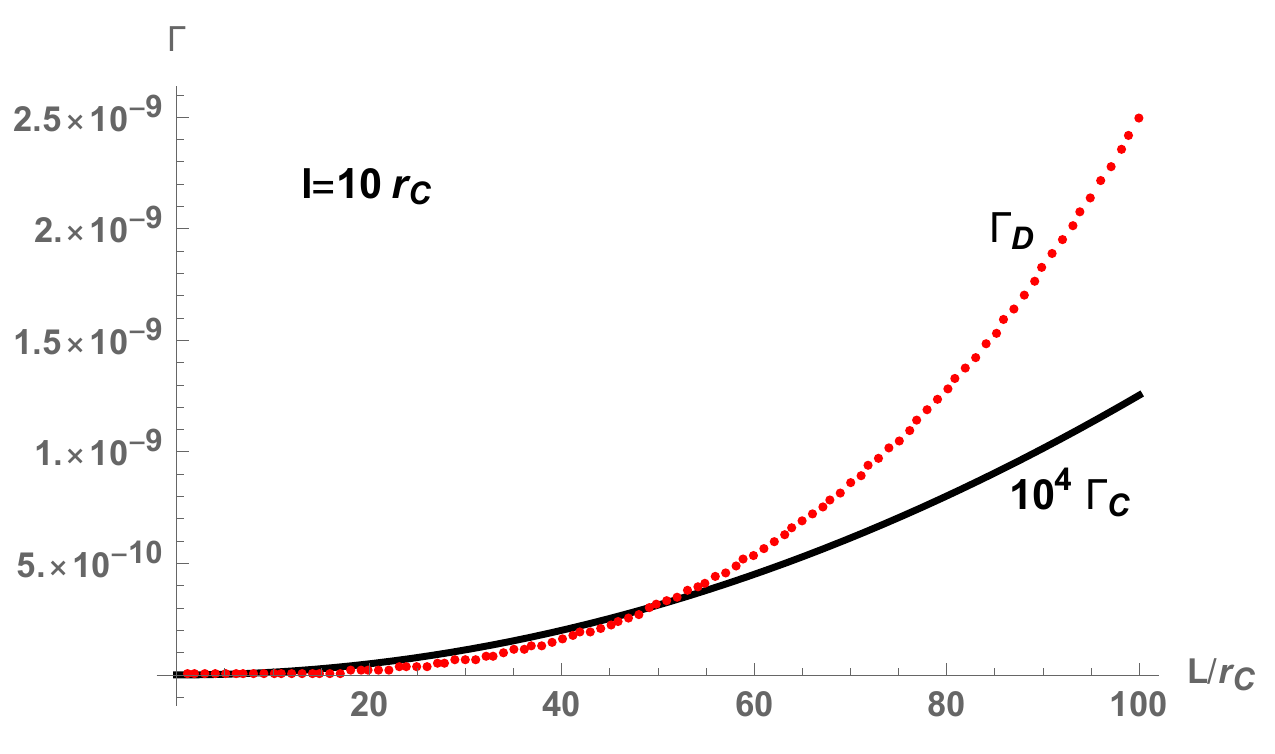}
  \end{minipage}
   \caption{Discrete ($\Gamma_{\text{\tiny{D}}}$, red dots) and continuous ($\Gamma_{\text{\tiny{C}}}$, black line) reduction rates for a cube as a function of its side $L$ in units of $r_C$. Left panel: $l=r_C$, $\varrho=10^{21}$ nucleons/m$^3$; right panel: $l=10\,r_C$, $\varrho=10^{18}$ nucleons/m$^3$. The other parameters are set as follows: $r_C=10^{-7}$ m, $\lambda=10^{-8}$ s$^{-1}$, $\Delta=10^{-3}\,r_C$.}
     \label{DvsC}
\end{figure}

Since Eq.~\eqref{gammaDser} involves doubles sums, we need to adapt  the EM formula to this case. We start by considering the double sum of a generic function of the difference of two variables, that satisfies the conditions of the previous theorem. By applying the EM formula to it (e.g. to the sum over $j$) we find:
\beq
\sum_{i,j=1}^Nf(i-j)=\sum_{i=1}^N\left(\int_0^Ndv\, f(i-v)+\frac{1}{2}\left[f(i-N)-f(i)\right]+\sum_{k=1}^{p}\frac{B_{2k}}{2k!}\left[\partial^{(2k-1)}_v f(i-v)\Big|^{v=N}_{v=0}\right]+R_p\right)\,,
\eeq
where $\partial^{(n)}_v$ denotes the $n$-th partial derivative with respect to the variable $v$. By applying the EM formula to $\sum_{i=1}^N\int_0^Ndv\, f(i-v)$ one can rewrite the previous equation as follows:
\beqa
\sum_{i,j=1}^Nf(i-j)&=&\int_0^Ndu \int_0^Ndv\, f(u-v)+\frac{1}{2}\left[f(0)-f(N)\right]+\sum_{k=1}^{p}\frac{B_{2k}}{2k!}\int_0^Ndv\left[\partial^{(2k-1)}_u f(u-v)\Big|^{u=N}_{u=0}\right]\nonumber\\
&&+\sum_{i=1}^N\sum_{k=1}^{p}\frac{B_{2k}}{2k!}\left[\partial^{(2k-1)}_v f(i-v)\Big|^{v=N}_{v=0}\right]+\tilde{R}_p\,,
\eeqa
where $\tilde{R}_p$ collects all the reminder terms. We now apply this equation to our case of interest, i.e. a function of the type $f(i-j)=\exp[-l^2(i-j)^2/4r_C^2]$. After some manipulation we obtain:
\beqa\label{EMerr}
\sum_{i,j=1}^Ne^{-\frac{l^2(i-j)^2}{4r_C^2}}&=&\int_0^Ndu \int_0^Ndv\, e^{-\frac{l^2(u-v)^2}{4r_C^2}}+\left(\frac{1}{2}-B_2\right)\left[1-e^{-\frac{l^2N^2}{4r_C^2}}\right]\nonumber\\
&&+2\sum_{k=1}^{p}\frac{B_{2k+2}}{(2k+2)!}\left[ \partial^{(2k)}_vf(N)-\partial^{(2k)}_vf(0)-\frac{1}{2}\partial^{(2k-1)}_vf(N)\right]+\tilde{R}_p\,,
\eeqa
where we have exploited the fact that $\partial^{(n)}_u f(u-v)=(-1)^n\,\partial^{(n)}_v f(u-v)$. Since the $n$-th derivative of a Gaussian function is an Hermite polynomial of order $n$ times the original Gaussian, one finds that the sum in the second line of Eq.~\eqref{EMerr} and the reminder $\tilde{R}_p$ are polynomials in $l^2/2r_C^2$, respectively of order $2p$ and $2p+1$. 
When $l\lesssim \sqrt{2}r_C$ the leading term of these polynomials is $l^2/2r_C^2$ 
and higher orders are negligible: whatever value of $p$ is chosen in Eq.~\eqref{EMerr}, the error is of the order $l^2/2r_C^2$. When $l\gtrsim \sqrt{2}r_C$ it is convenient to consider $p=0$ in Eq.~\eqref{EMerr} because this is the value of $p$ that minimizes the reminder $\tilde{R}_p$, which thus results of the order $l^2/2r_C^2$ (this is an example of when choosing larger $p$ does not improve the error estimate). 
We thus see that, for any value of the ratio $l/\sqrt{2}\,r_C$, the error made by neglecting the second line of  Eq.~\eqref{EMerr} is of the order $l^{2}/2r_C^{2}$. 
Accordingly, Eq.~\eqref{EMerr} can be rewritten as follows
\beq
\sum_{i,j=1}^Ne^{-\frac{l^2(i-j)^2}{4r_C^2}}=\int_0^Ndu \int_0^Ndv\, e^{-\frac{l^2(u-v)^2}{4r_C^2}}+\frac{1}{3}\left[1-e^{-\frac{L^2}{4r_C^2}}\right]+O\left(\frac{l^2}{2r_C^2}\right)\,,
\eeq
which with the help of Eq.~\eqref{gDelta} eventually allows to recover Eq.~\eqref{sumZ}. Figure~\ref{DvsC} shows the discrete ($\Gamma_{\text{\tiny{D}}}$, red dots) and continuous ($\Gamma_{\text{\tiny{C}}}$, black line) collapse rates for a cube as a function of its side $L$. In the left panel a lattice constant $l=r_C$ is taken, and as expected the difference between $\Gamma_{\text{\tiny{D}}}$ and $\Gamma_{\text{\tiny{C}}}$ is very small. For smaller values of $l$ the two lines are indistinguishable. The right panel shows $\Gamma_{\text{\tiny{D}}}$ and $\Gamma_{\text{\tiny{C}}}$ for a larger lattice constant $l=10\,r_C$: the two rates differ of about four orders of magnitude, thus showing that for $l>\sqrt{2}r_C$ the approximation given by $\Gamma_{\text{\tiny{C}}}$ is not good.

\end{document}